\documentclass[acmsmall]{acmart}

\AtBeginDocument{%
  \providecommand\BibTeX{{%
    \normalfont B\kern-0.5em{\scshape i\kern-0.25em b}\kern-0.8em\TeX}}}

\setcopyright{acmcopyright}
\copyrightyear{2018}
\acmYear{2018}
\acmDOI{10.1145/1122445.1122456}

\acmConference[Woodstock '18]{Woodstock '18: ACM Symposium on Neural
  Gaze Detection}{June 03--05, 2018}{Woodstock, NY}
\acmBooktitle{Woodstock '18: ACM Symposium on Neural Gaze Detection,
  June 03--05, 2018, Woodstock, NY}
\acmPrice{15.00}
\acmISBN{978-1-4503-XXXX-X/18/06}

\usepackage{csquotes}
\usepackage{listings, multicol}
\usepackage{xcolor}

\lstdefinelanguage{SQL}{
    keywords={TYPE, DATASET, CREATE, FEED, WITH, START, STOP, TO, CHANNEL, BROKER, INDEX, FUNCTION, LET, GROUP, BY, GROUP, AS, SELECT, WHERE, FROM, ORDER, DESC, LIMIT, REPETITIVE, SUBSCRIBE, ON, AT, USING, AND, PERIOD, ACTIVE, PRIMARY, KEY, CONTINUOUS, CONNECT}
}

\begin{document}

\title{Subscribing to Big Data at Scale}


\author{Xikui Wang}
\email{xikuiw@ics.uci.edu}
\affiliation{\institution{University of California Irvine}}

\author{Michael J. Carey}
\affiliation{\institution{University of California Irvine}}
\email{mjcarey@ics.uci.edu}

\author{Vassilis J. Tsotras}
\affiliation{\institution{University of California Riverside}}
\email{tsotras@cs.ucr.edu}


\begin{abstract}
Today, data is being actively generated by a variety of devices, services, and applications. Such data is important not only for the information that it contains, but also for its relationships to other data and to interested users. 
Most existing Big Data systems focus on \textbf{passively} answering queries from users, rather than \textbf{actively} collecting data, processing it, and serving it to users. 
To satisfy both passive and active requests at scale, users need either to heavily customize an existing passive Big Data system or to glue multiple systems together. Either choice would require significant effort from users and incur additional overhead. In this paper, we present the BAD (Big Active Data) system, which is designed to 
preserve the merits of passive Big Data systems and introduce new features for actively serving Big Data to users at scale. We show the design and implementation of the BAD system, demonstrate how BAD facilitates providing both passive and active data services, investigate the BAD system's performance at scale, and illustrate the complexities that would result from instead providing BAD-like services with a ``glued'' system.
\end{abstract}



\maketitle

\section{Introduction}
Big Data, without being analyzed, is merely a sequence of zeros and ones sitting on storage devices. To effectively utilize Big Data, researchers have developed a plethora of tools~\cite{shvachko2010hadoop, pig:paper, HivePaper, Spark}. 
In many applications today, we want not only to understand Big Data, but also to deliver subsets of interest proactively to interested users. In short, users should not only be able to \textbf{analyze data} but also to \textbf{subscribe to data}.
User subscription requests should not be limited to the incoming data's content but should also be able to consider its relationships to other data. 
Moreover, data to be sent should be allowed to include additional relevant and useful information.
We refer to this as the \textit{Big Active Data (BAD) challenge}. Due to the variety and volume of user requests, the data, and their relationships, analyzing, customizing, and delivering actionable data based on different user requests are not trivial tasks.

Traditionally, taking user requests and serving data continuously has been studied mostly in the context of Continuous Queries~\cite{tapestry, niagara_cq}. Users there register their requests as persistent queries and are subsequently notified whenever new results become available. Although the continuous query concept overlaps significantly with the active data problem, Big Data poses new challenges for classic continuous query approaches due to their complexity and computational costs. Similarly, triggers from traditional databases offer users the capability to react to events in a database under certain conditions~\cite{active_database_systems}. Users could try and take advantage of triggers to approach the active data challenge, but they soon become not applicable when the scale of the data, and thus the system, grows.

With the growth of streaming data and the need for real-time data analytics, Streaming Engines in recent years have been widely used in many active-data-related use cases~\cite{stratosphere, kreps2011kafka, spark_streaming_paper}. Data is ingested and optionally processed in streaming engines on-the-fly and then be pushed to other systems for later analysis. Streaming engines can be used for creating data processing and data customizing pipelines, but due to the nature of data streams, only a limited set of processing operations are available. As a result, streaming engines would
need to be coupled with other systems for meeting the complete BAD challenge at scale. This would introduce additional performance overhead and integration complexity for users.

Delivering data of interest to many users also resonates with the publish/subscribe communication paradigm from the distributed systems community~\cite{many_faces}. In the pub/sub paradigm, subscribers register their interests in incoming data items and will subsequently be notified about data published by publishers. Despite some similarity to the BAD challenge, pub/sub systems only forward data from publishers to subscribers without offering the capability to process it. Also, each data item is treated in isolation, so users' interests are limited to the data item itself (its topic, type, or content), but not its relationship to other data. In addition, pub/sub systems have to be integrated with other Big Data systems (e.g., Data warehouses) for supporting analytical queries.

One significant goal of the BAD approach advocated here is that users should not only be able to \textbf{analyze data} - i.e., to issue queries and receive result subsequently, but also to \textbf{subscribe to data} - i.e., to specify their interests in data and constantly receive the latest updates.
Many (passive) systems today support data analytics, but very few of them provide the \textbf{active features} we need. 
In addition to that, we would like to allow users to subscribe to data without always having to write independent queries. Mastering query languages could be useful for data analysts with expertise, but it might be a burden for end-users interested only in receiving data.
Although database features like stored procedures allow for the encapsulation of queries as executable units, they are still \textbf{passively} invoked by users. We need a system that allows users to analyze data declaratively and that enables users to subscribe to data actively with minimum effort.

In order to deliver the latest updates to end-users without asking them to construct queries and to ``pull'' data from the system constantly, we propose an abstraction - parameterized \textit{\textbf{data channels}} - to characterize user subscriptions.  
Users with expertise (e.g., application developers) can create data channels using declarative queries.
Users with interest in data (e.g., end-users)
can then subscribe to data channels with parameters and thus continuously receive new data. 
The BAD system runs data channels, manages their life-cycle, and offers them as active services.
This data channel abstraction provides a declarative user model for activating Big Data.

Previously, we implemented the initial prototype BAD system - BAD-RQ - on top of Apache AsterixDB~\cite{asterixdb14}. In BAD-RQ, we allow developers to create data channels using a declarative query language (SQL++) and enable users to subscribe to them by specifying their own parameters. Internally, channel queries are like parameterized prepared queries that are repetitively evaluated with subscription information and other relevant data. BAD-RQ computes them periodically on behalf of all users with all user-provided parameters and produces customized data for each subscribed user~\cite{jacobs2020bad}. 

As BAD-RQ executes channel queries periodically, users may attempt to leverage them to approximate continuous query semantics - obtaining updates incrementally without retrieving the entire history or reporting redundant results~\cite{tapestry}.
For example, a continuous query \textit{``send me \textbf{new} sensitive tweets''} can be loosely interpreted as a repetitive channel query \textit{``every 10 seconds, send me the sensitive tweets from the past 10 seconds"}.
Although users can approximate continuous query semantics with repetitive channels, BAD-RQ does not guarantee continuous semantics, and data items could be missed or redundantly reported. To ensure continuous semantics, we want a systematic way of supporting continuous queries in BAD. We need to make sure that users can receive incremental updates of data of interest with the guarantee of continuous semantics, to support different computational operations and indexes for accelerating evaluation, and to enhance the data channel model to provide a straightforward user model regarding continuous queries.

In this paper, we discuss BAD in-depth, present the BAD system, and introduce BAD-CQ - a new BAD service that provides continuous query semantics. We show how BAD-CQ is designed and implemented, and we investigate its performance under different workloads at scale. This paper is organized as follows: We review work related to BAD in Section 2. In Section 3, we dive into the detailed vision of Big Active Data, discuss the settings of the BAD problem, and describe the building blocks of a BAD system. In Section~\ref{sec:reptitive_bad}, we present a repetitive BAD use case to demonstrate the BAD-RQ service and illustrate the BAD user model. We introduce continuous BAD in Section~\ref{sec:continuous_bad}, discussing the limitations of approximating continuous BAD and presenting the design and implementation of the new BAD-CQ service. To compare a possible alternative approach with the BAD system, we introduce a GOOD (\textbf{G}luing \textbf{O}odles \textbf{O}f \textbf{D}ata platforms) system that consists of gluing together multiple Big Data systems in Section~\ref{sec:not_bad}. We show how to use the GOOD system for providing BAD services and illustrate the challenges that users would face in configuring, orchestrating, and managing such a glued system. We present a set of experimental results for the new BAD-CQ service and compare its performance with the glued system in Section~\ref{sec:exprs}.

\section{Related Work}

\textbf{Continuous Queries} are queries that are issued once and return results continuously as they become available. Tapestry~\cite{tapestry} first introduced Continuous Queries over append-only databases, defined continuous semantics, and created rewriting rules for transforming user-provided queries into incremental queries. Much subsequent research has focused on queries over streaming data. STREAM is a research prototype for processing continuous queries over data streams and stored data~\cite{stream}. It provides a Continuous Query Language (CQL) for constructing continuous queries against streams and updatable relations~\cite{stream_cql}. TelegraphCQ offers an adaptive continuous query engine that adjusts the processing during run-time and applies shared processing where possible~\cite{telegraph_cq}.
NiagaraCQ splits continuous queries into smaller queries and groups queries with the same expression signature together. It stores signature constants in a table and utilizes joins to evaluate grouped queries together to improve scalability, and it uses delta files for incremental evaluation on changed data to improve computational efficiency~\cite{niagara_cq}.
Most continuous query projects have been initial research prototypes, and very few of them have been scaled out to a distributed environment. This limits their applicability in Big Data use cases.

\textbf{Streaming Engines} allow low latency data processing and provide real-time analytics.
Apache Storm is a distributed stream processing framework. It provides two primitives, ``spouts'' and ``bolts'', to help users create topologies for processing data in real-time~\cite{storm}.
Spark Structured Streaming is a stream processing engine built on top of Apache Spark. It divides incoming data into micro-batches of Resilient Distributed Datasets (RDDs) for fault-tolerant stream processing, and it offers a declarative API for users to specify streaming computations~\cite{structred_streaming_paper, spark_streaming_paper}. 
Apache Kafka started as a distributed messaging system that allows collecting and delivering a high volume of log data with low latency. It later introduced a Streams API that enables users to create stream-processing applications~\cite{kreps2011kafka, link:kafka_streams}. 
Apache Flink~\cite{flink} (which originated from Stratosphere~\cite{stratosphere}) unifies both streaming and batch processing in one system and provides separate APIs (DataStream and DataSet) for creating programs running on a streaming dataflow engine~\cite{flink}. 
Due to the nature of streaming data, streaming engines usually do not store data for the long-term. The incoming data is processed and then soon pushed to other systems for further processing or persistence.

\textbf{Publish/subscribe Services} allow subscribers to register their interests in events and to be subsequently, asynchronously notified about events from publishers. There are three types of pub/sub schemes: topic-based, content-based, and type-based~\cite{many_faces}. 
In topic-based pub/sub, publication messages are associated with topics, and subscribers register their interests to receive messages about topics of interest. Many systems in this domain focus on providing scalable and robust pub/sub services, including Scribe~\cite{scribe}, SpiderCast~\cite{spidercast}, Magnet~\cite{magnet}, and Poldercast~\cite{poldercast}. 
Content-based pub/sub improves the expressiveness of pub/sub services by allowing subscriptions based on publications' content. Many research works in this area focus on improving the scalability and efficiency of matching publications to users' subscriptions, including XFilter~\cite{xfilter}, Siena~\cite{Siena}, YFilter~\cite{yfilter}, and BlueDove~\cite{blue_dove}. 
Type-based pub/sub groups publications based on their structure. It aims at integrating pub/sub services with (object-oriented) programming languages to improve performance~\cite{typebased_pubsub}.
While all these pub/sub services enable publishing data to a large number of subscribers, the expressiveness of subscriptions is limited as each publication is treated in isolation, and users often have to integrate a pub/sub service with other systems for data processing.

\section{Big Active Data}
To better
understand the Big Active Data (BAD) vision and the challenges in creating BAD services, in this section, we describe the BAD problem in detail, enumerate the requirements of a BAD system, and describe a set of BAD building blocks for fulfilling these requirements.

\subsection{A BAD World}
\label{sec:a_bad_world}
In a BAD world, data could come from various systems and services constantly and rapidly. Many users would like to acquire and share the data and use it for different purposes. 
Some users may want to analyze the collected incoming data for retrospective analysis. They may issue analytical queries like:

\begin{figure}[h]
    \centering
    \includegraphics[width=0.85\textwidth]{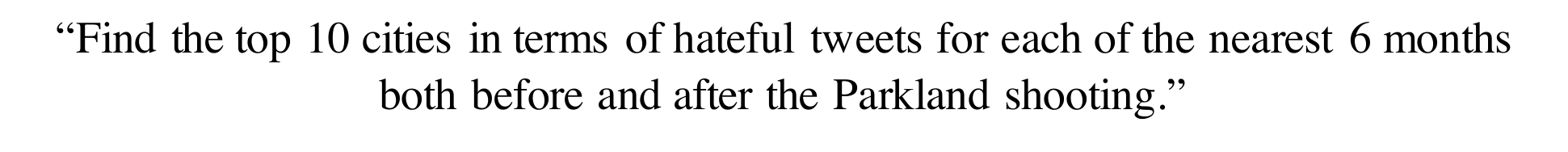}
    \caption{A sample analytical query on collected tweets}
    \label{fig:analytical_query}
\end{figure}

Other users may want to continuously receive updates regarding the data of interest to them. Users' interests may cover different aspects of the data. For example:

\begin{itemize}
    \item \textbf{Content}: Receive data when its content contains certain values - ``send me tweets that are hateful'';
    \item \textbf{Enrichment}: Receive data with relevant information - ``send me hateful tweets and their nearby schools'';
    \item \textbf{Relationship}: Receive data when it relates to other data - ``send me hateful tweets if they are near my location''.
\end{itemize}

Based on different needs of the users in the BAD world, we characterize three types of \textbf{BAD users}:

\begin{enumerate}
    \item \textbf{Analysts} issue queries to analyze collected incoming data and/or other relevant data.
    \item \textbf{Subscribers} make subscriptions and receive updates continuously using BAD applications. 
    \item \textbf{Developers} create BAD applications and provide BAD services to analysts and subscribers. 
\end{enumerate}

A full-fledged BAD system needs to serve all three types of users - analysts, subscribers, and developers – and should be able to scale to support a massive volume of data and a huge number of users.

\subsection{The BAD Building Blocks}
\label{sec:bad_blocks}
In order to provide the features described in Section~\ref{sec:a_bad_world}, a BAD system needs to have the following building blocks:

\begin{itemize}
    \item \textbf{Persistent Storage}: 
    In order to support retrospective analysis, data enrichment with relevant information, and customized data subscription, the BAD system should provide persistent storage to store collected incoming data, relevant data, and subscription information. It should be possible to add data to the BAD system through ingestion facilities, loading utilities, or applications' CRUD operations. Since data is persisted, developers should be able to utilize auxiliary data structures (like indexes) for accelerating data access.
    \item \textbf{Ingestion Facility}: 
    Data of interest, for either subscribers or analysts, may come into the BAD system rapidly. In order to capture such data, the BAD system should provide an ingestion facility to help continuously ingest data from various external data sources reliably and efficiently. BAD users should be able to easily create an ingestion pipeline in the BAD system without having to write low-level programs.
    \item \textbf{Analytical Engine}: 
    Data analytics enables analysts to reveal useful information from data. To help analysts understand the incoming data and its relationship with other relevant information, the BAD system should provide an analytical engine with support for declarative queries.
    \item \textbf{Data Channels}:
    In traditional Big Data applications, subscribers, who want to get data, rely on developers to translate their interest (subscriptions) into queries and then to retrieve data on behalf of subscribers. In practice, many subscriptions have similar structures like ``send me hateful tweets from city X'', ``send me hateful tweets near my location'', etc. To simplify creating BAD applications using the BAD system, we extract the shared structure among subscriptions and offer that as a service, namely a data channel, for subscribers to subscribe to with parameters. Data channels can be created using declarative queries and are managed by the BAD system.
    \item \textbf{Broker Network}: 
    Subscribers of a data channel expect the latest updates of their data of interest to be delivered to them continuously. The BAD system needs to handle millions of subscribers subscribing to a channel and to allow multiple channels to run concurrently. 
    Due to the volume of data exchanges between the BAD system and subscribers, the BAD system should include a broker network with caching and load-balancing strategies.
\end{itemize}

\begin{figure}[h]
    \centering
    \includegraphics[width=0.80\textwidth]{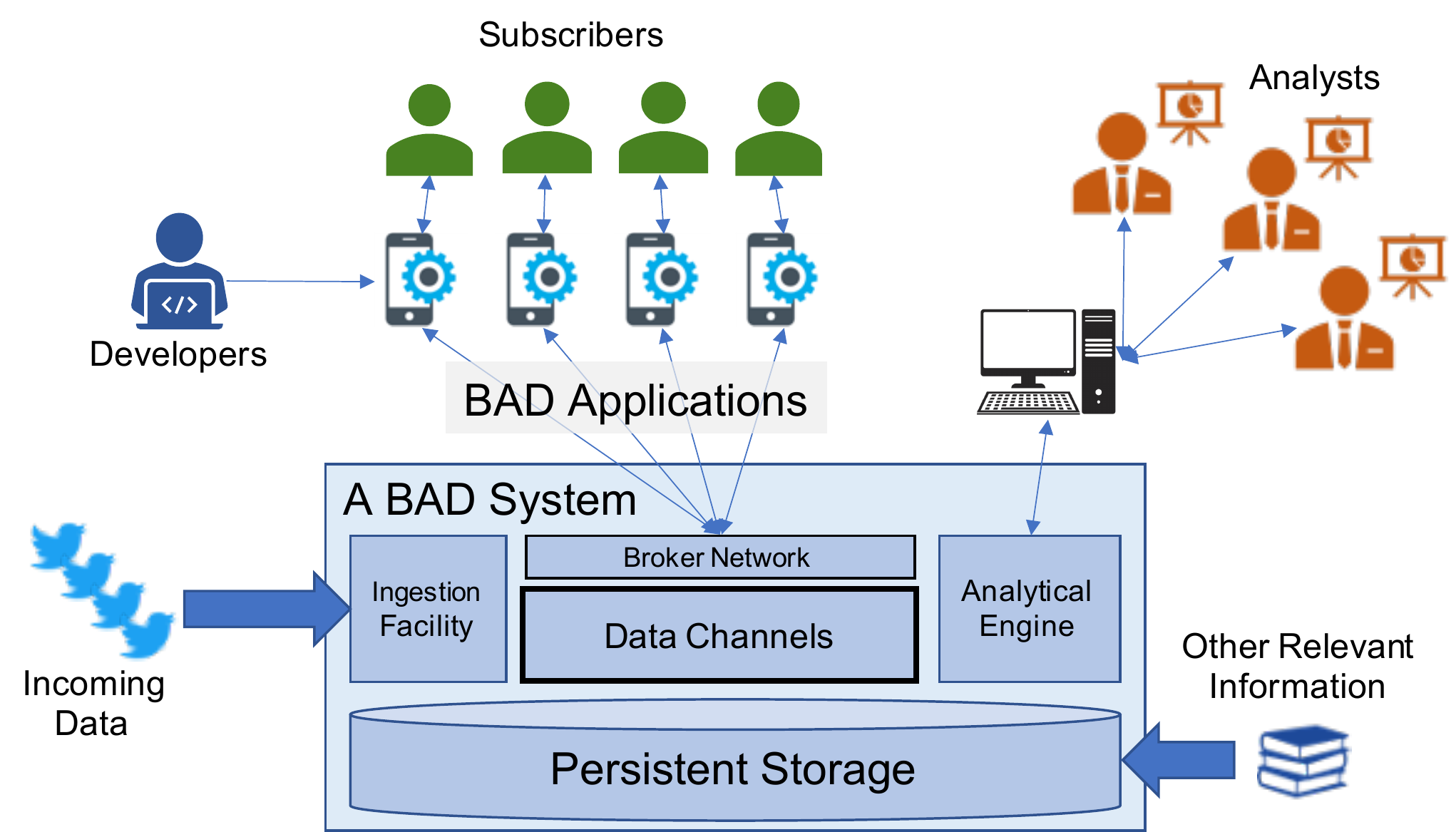}
    \caption{A BAD system for a BAD world}
    \label{fig:bad_system}
\end{figure}

We depict the BAD system and the BAD users it serves in Figure~\ref{fig:bad_system}. To the best of our knowledge, there is no existing Big Data platform that provides all the functionality needed from a BAD system. Some platforms can fulfill certain building blocks in the BAD system, but one would have to hand-wire multiple systems together to get all desired BAD features. 
A well-designed, integrated, and efficient BAD system with support for a declarative language can significantly reduce the effort required to create BAD services.

\section{Repetitive BAD: BAD-RQ}
\label{sec:reptitive_bad}
A straightforward way of creating a BAD system is to 
``activate'' an existing Big Data system by adding the missing building blocks needed for BAD services and making sure it scales to millions of subscribers. We created the initial prototype BAD system - BAD-RQ - on top of Apache AsterixDB, an open-source Big Data Management System that provides distributed data management for large-scale, semi-structured data. In this section, we focus on the user model of BAD-RQ and illustrate how developers utilize it to create BAD services. Interested readers can refer to \cite{breakingbad, jacobs2017bad, bad_broker, jacobs2020bad, bad_to_the_bone} for a more detailed description of the whole BAD project.

\subsection{A BAD Repetitive Use Case}
To illustrate BAD-RQ, we use a sample scenario in which we want to provide BAD services to police officers around tweets. Users of these services include investigative officers as \textbf{analysts} who want to study tweets about certain events, and in-field officers as \textbf{subscribers} who patrol around the city and want to receive live tweets meeting certain requirements. 
Tweets come into BAD-RQ from an external system continuously, and each contains a hateful flag provided by the datasource indicating whether this tweet is hateful and may relate to a potential crime. Location updates of patrolling in-field officers are also sent to BAD-RQ constantly to show their latest location. We describe the implementation of BAD building blocks in BAD-RQ and demonstrate how \textbf{developers} can utilize them for creating BAD services.

\subsubsection{Persistent Storage}
\label{sec:persistent_storage}
In order to support queries from analysts and subscriptions from subscribers, both incoming tweets and location updates need to be persisted in the BAD system. BAD-RQ offers the same storage functionality as AsterixDB, including all data types and indexes. AsterixDB organizes data under \textit{dataverses} (similar to databases in an RDBMS). Without loss of generality, all data discussed in this section is stored in the ``BAD'' dataverse. 

To store data in the BAD dataverse, we (as developers) need to create a \textit{datatype}, which describes the stored data, and a \textit{dataset}, which is a collection of records of a datatype. We define both the Tweet and OfficerLocation data types as ``open'', which makes the stored data extensible. The ``hateful\_flag'' attribute, indicating whether a tweet is hateful, is not specified in the data type and thus it is an open (optional) field. 
When ``hateful\_flag'' is not provided but needed for a BAD application, a developer could use a enrichment user-defined function (UDF) to enrich tweets during data ingestion~\cite{xikui_idea}. We create a dataset \textit{Tweets} for storing incoming tweets, a dataset \textit{OfficerLocations} for storing location updates, and two R-Tree indexes on the location attribute of each dataset for more efficient data access. The DDL statements for creating both datasets are shown in Figures~\ref{ddl:tweets} and \ref{ddl:officer} respectively.

\begin{figure}[h]
\footnotesize
\centering
\begin{lstlisting}[
           language=SQL,
           basicstyle=\ttfamily,
           showstringspaces=false,
           commentstyle=\color{gray}
        ]
  CREATE TYPE Tweet AS OPEN {
    tid: bigint,
    area_code: string,
    location: point
  };
  CREATE DATASET Tweets(Tweet) PRIMARY KEY tid;
  CREATE INDEX t_location ON Tweets(location) TYPE RTREE;
\end{lstlisting}
\caption{Datatype and dataset definition for Tweets}
\label{ddl:tweets}
\end{figure}

\begin{figure}[h]
\footnotesize
\centering
\begin{lstlisting}[
           language=SQL,
           basicstyle=\ttfamily,
           showstringspaces=false,
           commentstyle=\color{gray}
        ]
  CREATE TYPE OfficerLocation AS OPEN {
    oid: int,
    location: point
  };
  CREATE DATASET OfficerLocations(OfficerLocation) PRIMARY KEY oid;
  CREATE INDEX o_location ON OfficerLocations(location) TYPE RTREE;
\end{lstlisting}
\caption{Datatype and dataset definition for officer location updates}
\label{ddl:officer}
\end{figure}

\subsubsection{Ingestion Facility}
\label{sec:ing_facility}
Since tweets and location updates may come at a very rapid rate, the BAD system needs to intake such ``fast'' incoming data efficiently. AsterixDB provides data feeds for data ingestion from various data sources with different data formats~\cite{raman_feeds}. We create a socket data feed to intake JSON formatted tweets using the statements shown in Figure~\ref{ddl:feed_tweet}.
Similarly, we create a data feed for intaking location updates sent by in-field officers in Figure~\ref{ddl:feed_officer}. 
In this use case, we send in-field officers nearby hateful tweets based only on their current location, so we create an UPSERT (i.e., insert if new, else replace) data feed by setting \textit{``insert-feed''} to false. 
In cases where officers' entire movement history is needed, one can also create an INSERT data feed like the one used for tweets.

\begin{figure}
\footnotesize
\begin{lstlisting}[
           language=SQL,
           basicstyle=\ttfamily,
           showstringspaces=false,
           commentstyle=\color{gray}
        ]    
  CREATE FEED TweetFeed WITH {
    "type-name" : "Tweet",
    "adapter-name": "socket_adapter",
    "format" : "JSON",
    "sockets": "127.0.0.1:10001",
    "address-type": "IP",
    "insert-feed" : true
  };
  CONNECT FEED TweetFeed TO DATASET Tweets;
  START FEED TweetFeed;
\end{lstlisting}
\caption{A data feed for ingesting tweets}
\label{ddl:feed_tweet}
\end{figure}

\begin{figure}
\footnotesize
\begin{lstlisting}[
           language=SQL,
           basicstyle=\ttfamily,
           showstringspaces=false,
           commentstyle=\color{gray}
        ]    
  CREATE FEED LocationFeed WITH {
    "type-name" : "OfficerLocation",
    "adapter-name": "socket_adapter",
    "format" : "JSON",
    "sockets": "127.0.0.1:10002",
    "address-type": "IP",
    "insert-feed" : false
  };
  CONNECT FEED LocationFeed TO DATASET OfficerLocations;
  START FEED LocationFeed;
\end{lstlisting}
\caption{A data feed for ingesting location updates}
\label{ddl:feed_officer}
\end{figure}

\subsubsection{Analytical Engine}
BAD-RQ supports data analytics using the query engine in AsterixDB. It provides SQL++~\cite{sql++don, ong2014sql++} (a SQL-inspired query language for semi-structured data) for users to construct analytical queries. SQL++ supports standard SQL query operations (SELECT, JOIN, GROUP BY, ORDER BY, etc.), spatial-temporal queries, operations designed for semi-structured data, etc. One can use the SQL++ query shown in Figure~\ref{query:analytics} to answer the analytical query from Figure~\ref{fig:analytical_query}. For a query executed multiple times with different constant expressions, analysts can also define it as a SQL++ UDF and invoke it with parameters instead of re-constructing the same query every time. As an example, the analytical query in Figure~\ref{query:analytics} can be encapsulated in the SQL++ UDF shown in Figure~\ref{query:top_city_udf}.

\begin{figure}
\scriptsize
\begin{lstlisting}[
           language=SQL,
           basicstyle=\ttfamily,
           showstringspaces=false,
           commentstyle=\color{gray}
        ]    
  LET stime = datetime("2017-07-14T10:10:00"), etime = datetime("2018-08-14T10:10:00")
  FROM Tweets t WHERE t.timestamp > stime AND t.timestamp < etime
  GROUP BY print_datetime(t.timestamp, "Y-M")
  GROUP AS TweetsByMonth
  SELECT print_datetime(t.timestamp, "Y-M") AS Month, (
  	SELECT VALUE tbm.t.area_code FROM TweetsByMonth tbm 
  	GROUP BY tbm.t.area_code ORDER BY count(1) DESC LIMIT 10
  ) MostHatefulCities;
\end{lstlisting}
\caption{An SQL++ query looking for the 10 most hateful cities in each month in a given time frame}
\label{query:analytics}
\end{figure}

\begin{figure}
\scriptsize
\begin{lstlisting}[
           language=SQL,
           basicstyle=\ttfamily,
           showstringspaces=false,
           commentstyle=\color{gray}
        ]    
CREATE FUNCTION mostHatefulCitiesByMonth(stime,etime) {
  FROM Tweets t WHERE t.timestamp > stime AND t.timestamp < etime
  GROUP BY print_datetime(t.timestamp, "Y-M")
  GROUP AS TweetsByMonth
  SELECT print_datetime(t.timestamp, "Y-M") AS Month, (
  	SELECT VALUE tbm.t.area_code FROM TweetsByMonth tbm 
  	GROUP BY tbm.t.area_code ORDER BY count(1) DESC LIMIT 10
  ) MostHatefulCities
};
mostHatefulCitiesByMonth(datetime("2017-07-14T10:10:00"),datetime("2018-08-14T10:10:00"));
\end{lstlisting}
\caption{A UDF based on an analytical query}
\label{query:top_city_udf}
\end{figure}

\subsubsection{Data Channels}
\label{sec:rep_chn_definition}
Since queries can be encapsulated as a UDF with parameters, and subscriptions with a similar structure can also be interpreted as a parameterized query, 
we can use a SQL++ UDF to group these subscriptions together and ``activate'' it as a data channel.
Developers can create data channels based on SQL++ UDFs and offer them as services, and subscribers can subscribe to them with parameters to receive data of interest subsequently.
As an example, if in-field officers want to know the number of hateful tweets near their current location in the past hour, we can first create the UDF in Figure~\ref{ddl:recent_hateful_tweets_count_udf}, which can be invoked using an officer's ID and returns the number of recent hateful tweets nearby.
We ``activate'' this UDF using the statement in Figure~\ref{ddl:create_recent_hateful_tweets_channel} by creating a data channel using this UDF. This channel has a configurable period ``10 minutes'' indicating that it computes every 10 minutes. In-field officers who subscribed to this channel then will receive the number of nearby hateful tweets in the past hour every 10 minutes. 
We will further discuss how a channel evaluation produces customized data for each subscriber in Section~\ref{sec:channel_computing_model}.

\begin{figure}[h]
\footnotesize
\begin{lstlisting}[
           language=SQL,
           basicstyle=\ttfamily,
           showstringspaces=false,
           commentstyle=\color{gray}
        ]    
CREATE FUNCTION RecentNearbyHatefulTweetsCount(oid) {
  FROM OfficerLocations o, Tweets t
  WHERE o.oid = oid AND t.hateful_flag = true
    AND spatial_distance(t.location, o.location) < 5 
    AND t.timestamp > current_datetime() - day_time_duration("PT1H")
  SELECT count(*) AS HatefulTweetsNum, current_datetime() AS CurrentTime
};
RecentNearbyHatefulTweetsCount("0907");
\end{lstlisting}
\caption{An UDF for counting hateful tweets near certain in-field officer given his/her officer ID}
\label{ddl:recent_hateful_tweets_count_udf}
\end{figure}

\begin{figure}[h]
\footnotesize
\begin{lstlisting}[
           language=SQL,
           basicstyle=\ttfamily,
           showstringspaces=false,
           commentstyle=\color{gray}
        ]    
  CREATE REPETITIVE CHANNEL RecentNearbyHatefulTweetCountChannel 
    USING RecentNearbyHatefulTweetsCount@1 PERIOD duration("PT10M");
\end{lstlisting}
\caption{Creating a data channel based on a UDF with a parameter}
\label{ddl:create_recent_hateful_tweets_channel}
\end{figure}

\subsubsection{Brokers and Subscriptions}
The BAD system includes a broker sub-system for managing the communication with a large number of subscribers. A broker could be a single server that only forwards customized data to subscribers or a broker network that provides load balancing, subscription migration, and different caching strategies. Interested readers can refer to \cite{bad_broker, hang_broker} for more details. A developer can choose a broker suited for the use case and register it as an HTTP endpoint in the BAD system as in Figure~\ref{ddl:create_broker}. A subscriber can then subscribe to a channel in the BAD system on this broker using the statement in Figure~\ref{ddl:subscribe}. 
A given channel execution can produce customized data for subscribers subscribed on different brokers, and the customized data is sent to the corresponding brokers based on which brokers the subscriptions are subscribed on. A broker receives the customized data from channel executions and then disseminates it to its subscribers.

\begin{figure}[h]
\footnotesize
\begin{lstlisting}[
           language=SQL,
           basicstyle=\ttfamily,
           showstringspaces=false,
           commentstyle=\color{gray}
        ]    
  CREATE BROKER BROKER_A AT "http://BROKER_A_HOST:BROKER_A_PORT/API";
\end{lstlisting}
\caption{Registering a broker to BAD}
\label{ddl:create_broker}
\end{figure}

\begin{figure}[h]
\footnotesize
\begin{lstlisting}[
           language=SQL,
           basicstyle=\ttfamily,
           showstringspaces=false,
           commentstyle=\color{gray}
        ]
  SUBSCRIBE TO RecentNearbyHatefulTweetCountChannel("0907") ON BROKER_A;
\end{lstlisting}
\caption{Subscribing to a channel with parameters on a broker}
\label{ddl:subscribe}
\end{figure}

\subsection{Data Channel Evaluation}
\label{sec:channel_computing_model}
As the core feature of the BAD system, data channels combine incoming data, relevant information, subscriptions, and broker information to produce customized data for each subscriber. In this section, we describe how BAD-RQ evaluates data channels to support a large number of subscriptions at scale.

\subsubsection{Modeling Brokers and Subscriptions}
As we mentioned in Section~\ref{sec:bad_blocks}, subscribers subscribe to a data channel with parameters, and there could be millions of subscribers for a data channel. Given the large volume of subscriptions, separately evaluating a channel query (the underlying UDF of a channel) for each subscriber would be too computationally expensive. Inspired by \cite{niagara_cq}, BAD-RQ
stores subscriptions as data and evaluates the channel query using the analytical query engine. Benefiting from the query optimization, indexes, and distributed evaluation in AsterixDB, BAD-RQ can compute a channel query with a lot of subscriptions efficiently, and the channel evaluation process can also take advantage of the shared computation among subscriptions in order to serve more subscribers.

BAD-RQ uses the data types defined in Figure~\ref{ddl:broker_sub_definition} to store the broker and subscription information internally. 
Broker information is decoupled from subscriptions, so a broker record can be modified without having to update all related subscriptions. 
The subscription data type is defined as \textit{open}, and the parameters of a subscription are stored as open attributes and named as param0, param1, etc. This allows a data channel to support an arbitrary number of parameters with arbitrary data types. 
The broker dataset is a permanent part of the BAD-RQ metadata.
The subscription dataset is tied to the life-cycle of a data channel.
When a developer creates a data channel (e.g., RecentNearbyHatefulTweetCountChannel), a corresponding subscription dataset (RecentNearbyHatefulTweetCountChannelSubscriptions) is also created, and this will be removed when the channel is dropped. Whenever a subscriber subscribes to the channel, a new subscription record is inserted into the subscription dataset.
 
\begin{figure}[h]
\footnotesize
\begin{lstlisting}[
           language=SQL,
           basicstyle=\ttfamily,
           showstringspaces=false,
           commentstyle=\color{gray}
        ]
    CREATE TYPE Brokers AS {
        dataverse_name: string,
        broker_name: string,
        broker_end_point: string
    };
    CREATE TYPE Subscription AS {
        subscription_id: uuid,
        broker_name: string,
        dataverse_name: string
    };
\end{lstlisting}
\caption{Data type definitions for brokers and subscriptions (internal to BAD)}
\label{ddl:broker_sub_definition}
\end{figure}

\subsubsection{An example of Channel Evaluation} 
\begin{figure}[h]
    \centering
    \includegraphics[width=0.95\textwidth]{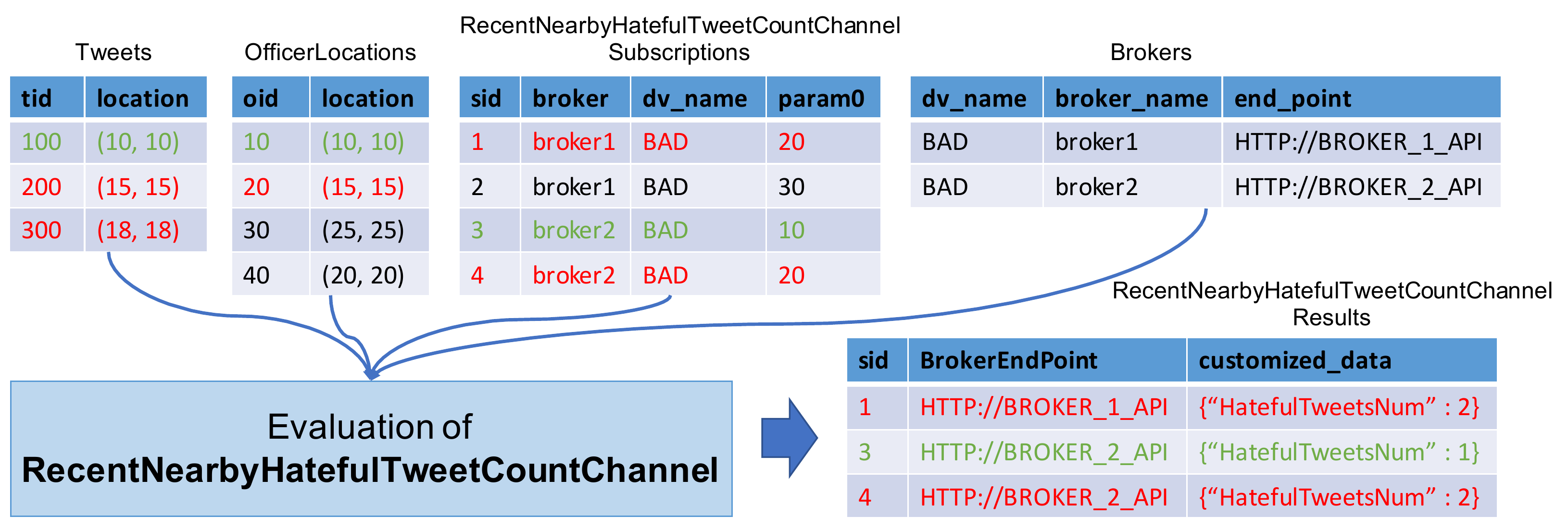}
    \caption{A data sample for evaluating a data channel}
    \label{fig:channel_io}
\end{figure}

In order to illustrate how BAD-RQ periodically computes a channel and produces customized data for each subscriber using broker and subscription information, we show a small data sample in Figure~\ref{fig:channel_io} for the channel defined in Section~\ref{sec:rep_chn_definition}, which returns the number of hateful tweets near a particular in-field officer. 
For illustrative simplicity, we assume all three tweets are posted within one hour and are hateful, and attributes not used for evaluation are not shown in the figure.
The channel evaluation combines information from four datasets, including \textit{OfficerLocations}, \textit{Tweets}, \textit{RecentNearbyHatefulTweetCountChannelSubscriptions}, and \textit{Brokers}, and it produces the customized data shown in the \textit{RecentNearbyHatefulTweetCountChannelResults} dataset. Related tuples are colored the same. 
Taking red tuples as an example, we find two tweets near officer with oid 20's current location at (15, 15): tweet 200 at (15, 15) and tweet 300 at (18, 18). Also, there are two subscriptions (subscription 1 and subscription 4) subscribe to the nearby hateful tweet number of officer 20 (having param0 equal to 20). Subscription 1 is on broker 1, and subscription 4 is on broker 2. Based on the above information, BAD produces two notifications, one for each subscriber, and sends them to their corresponding broker APIs.

\subsubsection{Channel Evaluation Internals} 
\begin{figure}[h]
    \centering
    \includegraphics[width=0.85\textwidth]{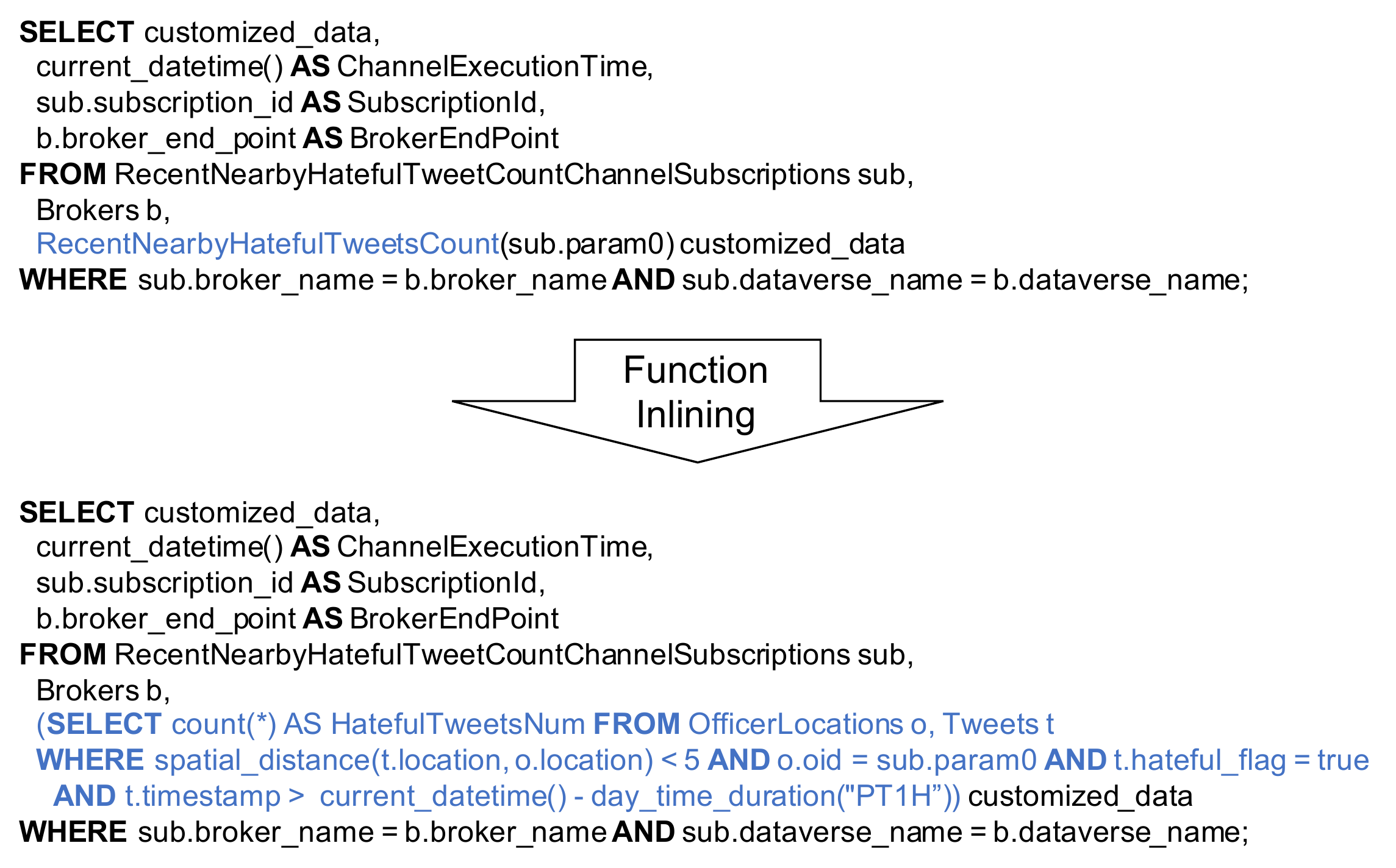}
    \caption{An illustrative query for computing a channel}
    \label{ddl:channel_evaluation}
\end{figure}

Evaluating a channel is equivalent to evaluating a query where we apply the underlying UDF to parameters from subscriptions to produce customized data. For example, evaluating the channel defined back in Figure~\ref{ddl:create_recent_hateful_tweets_channel} is equivalent to evaluating the query shown in Figure~\ref{ddl:channel_evaluation}. In this query, we apply the UDF in Figure~\ref{ddl:recent_hateful_tweets_count_udf} on parameters from subscriptions and nest the return value of the UDF into a ``customized\_data" field. The UDF can be inlined into the query, as shown, and be compiled and optimized together with it. The broker endpoint and subscription ID are also attached to each customized data record. The broker endpoint is used for the channel to send the result to a corresponding broker API, and the subscription ID is used by brokers to identify which subscriber the customized data should be delivered to.

\begin{figure}[h]
    \centering
    \includegraphics[width=0.90\textwidth]{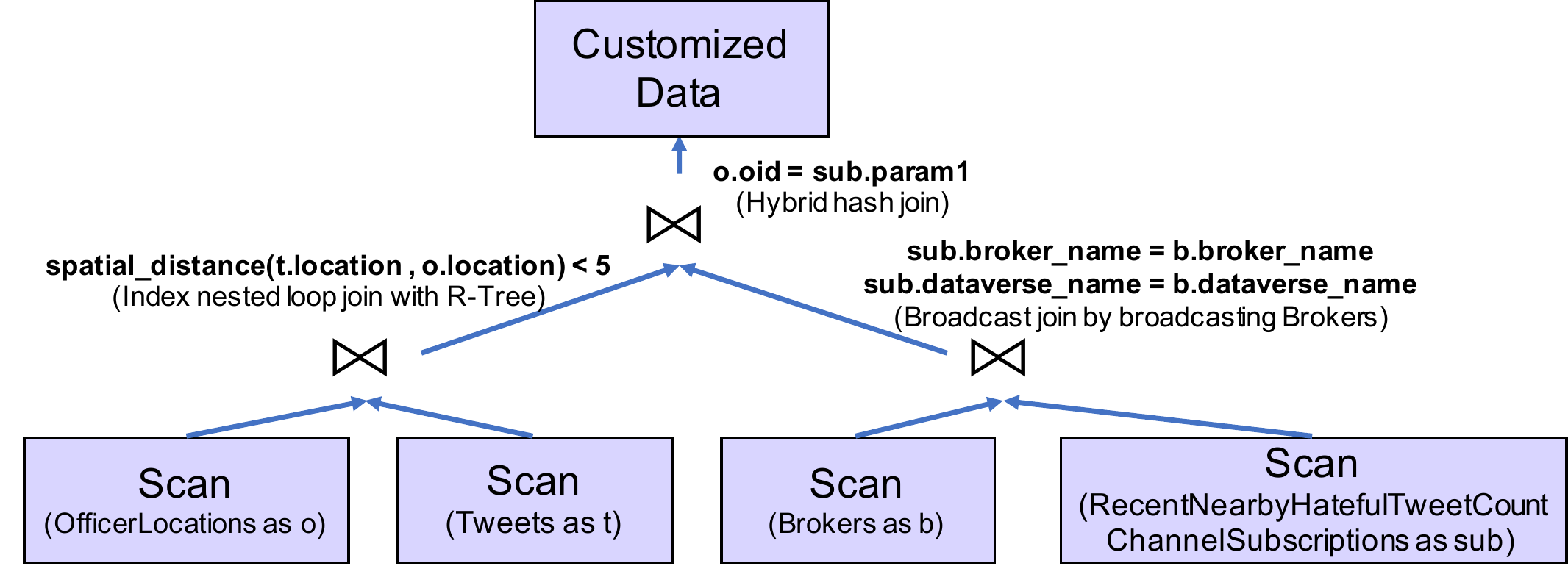}
    \caption{A query plan for channel evaluation}
    \label{fig:channel_plan}
\end{figure}

Since the query used for evaluating a channel is computed on the analytical engine of AsterixDB, it can be optimized by the query optimizer and be accelerated by utilizing efficient algorithms and indexes. Under the hood, the query in Figure~\ref{ddl:channel_evaluation} compiles into a query plan as shown in Figure~\ref{fig:channel_plan}. BAD can use use an R-Tree index to accelerate the spatial join between \textit{Tweets} and \textit{OfficerLocations}. Also, since the number of brokers is small compared with subscriptions, it can broadcast the Brokers to avoid unnecessary shuffling of the Subscriptions dataset. It can use a hybrid hash join to join the two intermediate results in parallel on all nodes in the cluster.

\subsubsection{Customized Data Delivery}
\label{sec:bad_data_delivery} A data channel executes on a specified period (time interval) to generate customized data. Depending on subscribers' preferences, the customized data can either be eagerly or lazily delivered. In the eager (push) mode, the produced data is pushed to brokers directly so they can immediately disseminate the data to subscribers. As the produced subscription result data is not persisted in BAD-RQ in this mode, brokers have to be fault-tolerant to avoid data loss. In the lazy (pull) mode, the customized data is first persisted in the BAD-RQ storage engine. The channel then sends a notification to the brokers whose subscribers have customized data that was produced in this channel execution. A broker that receives such a notification then pulls the customized data from BAD-RQ and distributes it to the subscribers. To this end, a result dataset (RecentNearbyHatefulTweetCountChannelResults) is created for persisting produced customized data when a ``lazy'' channel is created. The result dataset has an index on the ``ChannelExecutionTime'' attribute for accelerating result pulling. 
Since the customized data is persisted in the storage engine in this mode, brokers then have the flexibility to choose when to disseminate the notifications to subscribers, and the storage engine ensures data safety. BAD-RQ uses the pull (broker-initiated) mode as the default mode for its channels.

\section{Continuous BAD: BAD-CQ}
\label{sec:continuous_bad}
BAD-RQ ``activates'' a UDF (a parameterized query) to create a data channel that allows subscribers to constantly receive updates of interesting data. 
Although BAD-RQ demonstrates how to transform a ``passive'' Big Data system into a basic ``active'' one for creating BAD services, it faces several limitations when users have more requirements. 

In some use cases, subscribers may want the latest information delivered \textbf{incrementally}. Examples include ``send me new hateful tweets on campus'', ``notify me when an emergency happens around me'', and ``let me know when crimes happen near my house''. We call such use cases \textbf{Continuous BAD}. In order to support them, data channels in BAD need to provide continuous semantics, in which they continuously return incremental updates. Developers using BAD-RQ could try to approximate continuous semantics using repetitive channels, but such approximations would face challenges due to the lack of native support for true continuous semantics. In this section, we look at an example of continuous BAD and demonstrate how to use BAD-RQ to approximate it. We discuss the limitations of this approximation and then introduce a new BAD service - BAD-CQ - designed for supporting continuous BAD.

\subsection{Approximately Continuous Queries}
\label{sec:approx_c}
To illustrate continuous BAD and its BAD-RQ approximation, we look at a simple continuous use case where \textit{``in-field officers (subscribers) want to know \textbf{new} hateful tweets near their current location''}. We introduce the setup for approximating continuous semantics in BAD-RQ and show how to construct a repetitive channel query for this approximation.

\subsubsection{BAD Timestamps}
\label{sec:bad_timestmaps}
As subscribers are interested in \textbf{new} tweets, BAD-RQ needs to determine which portion of the collected tweets are new (i.e., tweets ingested but not yet reported). Different from streaming engines where all data in the engine is \textbf{new}, and \textbf{old} data is aged out, BAD-RQ keeps all data in the storage for supporting other services (e.g., data analytics). In order to differentiate new data from old, BAD-RQ needs to utilize timestamps.

In some cases (like tweets), incoming data comes with a ``timestamp'' attribute which indicates when was a data item created (a.k.a., valid time~\cite{temporal_db} or event time~\cite{flink}). 
This attribute could potentially be used for differentiating new tweets from old ones. 
However, this would introduce additional complexity in handling out-of-order arrivals.
Besides, when such an attribute is not provided in the incoming data, we still need to find another solution\footnote{Streaming Engines (such as Spark Structured Streaming) that compute with event time offer watermarking to handle late arrivals. BAD-RQ with BAD timestamps (and BAD\_CQ later introduced in  Section~\ref{sec:bad-cq}) can provide similar functionality with proper channel queries. Here we focus on the general use cases without assuming the existence of event time.}.
BAD-RQ allows developers to attach timestamps to incoming tuples during data ingestion by attaching a UDF to the ingestion pipeline. For this use case, we can create the UDF shown in Figure~\ref{ddl:ing_udf} and attach it to the tweet data feed defined in Figure~\ref{ddl:feed_tweet}. This UDF adds an ``ingested\_timestamp'' attribute to each incoming tweet, which marks the current date time when the tweet first enters the pipeline (a.k.a, ingestion time~\cite{flink}). We can utilize this timestamp to determine whether or not a tweet is \textbf{new}.

\begin{figure}[h]
\footnotesize
\begin{lstlisting}[
           language=SQL,
           basicstyle=\ttfamily,
           showstringspaces=false,
           commentstyle=\color{gray}
        ]
  CREATE FUNCTION AddIngestionTime(incoming_record) {
    object_merge({"ingested_timestamp": current_datetime()}, incoming_record)
  };
\end{lstlisting}
\caption{A UDF for adding ingestion time}
\label{ddl:ing_udf}
\end{figure}

\subsubsection{A Repetitive Approximation}
\label{sec:a_repetitive_approc}
By assigning timestamps to incoming tweets during ingestion, we can use the ingestion timestamp to infer the arrival order of tweets and differentiate ``new'' tweets from ``old''. We can construct a repetitive data channel with a designated channel period, as shown in Figure~\ref{ddl:new_nearby_hateful_tweets}. In this channel, we look for hateful tweets ingested in the past 10 seconds from the time when the channel executes.
These tweets are new and thus haven't been examined yet. We join them with officers' current locations and look for nearby new hateful tweets for each subscribed in-field officer. The channel is defined to execute every 10 seconds, so subscribers can continuously receive new nearby hateful tweets. This allows us to approximate continuous (incremental) semantics with a repetitively executed channel query that runs every 10 seconds and looks back 10 seconds.

\begin{figure}[h]
\scriptsize
\begin{lstlisting}[
           language=SQL,
           basicstyle=\ttfamily,
           showstringspaces=false,
           commentstyle=\color{gray}
        ]
CREATE FUNCTION NewNearbyHatefulTweets(oid){
  SELECT t
  FROM OfficerLocations o, Tweets t
  WHERE spatial_distance(t.location, o.location) < 5 AND o.oid = oid AND t.hateful_flag = true
    AND t.ingested_timestamp > current_datetime() - day_time_duration("PT10S")
};
CREATE REPETITIVE CHANNEL NewNearbyHatefulTweetsChannel USING NewNearbyHatefulTweets@1 
PERIOD duration("PT10S");
\end{lstlisting}
\caption{A repetitive data channel looking for new nearby hateful tweets}
\label{ddl:new_nearby_hateful_tweets}
\end{figure}

\subsubsection{Challenges in Approximation}
Although developers could use BAD-RQ to approximate continuous semantics just as shown, 
such an approximation is not perfect in practice and could fail to have continuous semantics in some circumstances. Also, due to the lack of native syntax support for continuous semantics, constructing an approximation query can become very complex. Challenges include:

\begin{itemize}
    \item \textit{Scheduling Delay:} We approximate the continuous semantics by examining data ingested in the past execution period (e.g., 10 seconds) from the current channel execution time. To perfectly approximate a continuous query, we rely on BAD-RQ to schedule the channel execution \textit{on time} to make sure that all incoming data is examined. However, this is  impractical in practice, especially in a distributed environment. 
    If a scheduling delay happens, some data can be missed by the channel, as shown in Figure~\ref{fig:missing_data}. 
    This channel executes every 10 seconds and examines data ingested from the past 10 seconds. If the actual channel execution 1 is delayed from T = 20 to 20.5, the data ingested from T = 10 to 10.5 will not be examined and thus missed.
    \begin{figure}[h]
        \centering
        \includegraphics[width=0.90\textwidth]{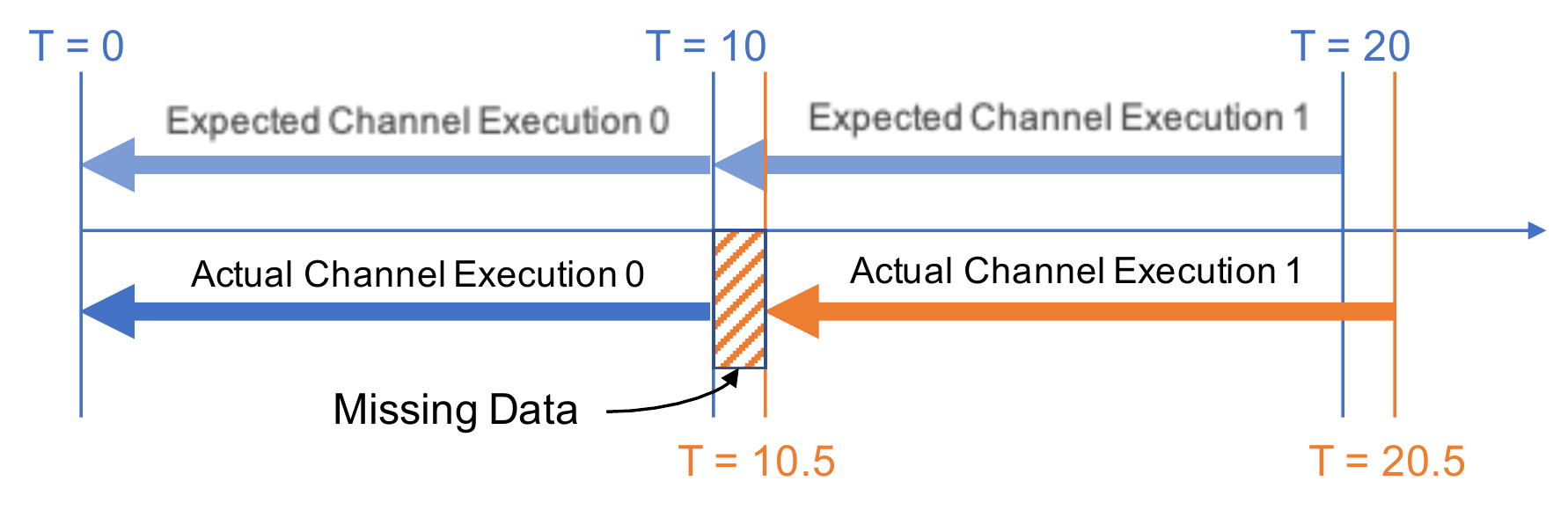}
        \caption{Missing data due to scheduling delays}
        \label{fig:missing_data}
    \end{figure}
    \item \textit{Early Timestamping:} 
    The approximation of BAD-RQ uses the ingestion timestamp for determining whether ingested data should be examined in a channel execution. However, since the ingested (timestamped) data does not become visible to channel execution instantaneously due to delays in data transmission, data enrichment (if any), secondary index(es) updating (if any), primary index updating, and waiting for the storage transaction to complete, there is a chance that a running channel execution could miss the data just ingested, even if the channel execution is scheduled on time. 
    This is illustrated in Figure~\ref{fig:early_timestamping}, where channel execution 1 starts at $T = 10$ and a tuple t100 is ingested at $T = 10 - \sigma$ and later persisted and becomes visible to queries at $T = 10 + \delta$ due to the delay~\footnote{In practice, this time gap is very small. We emphasize the delay in Figure~\ref{fig:early_timestamping} for illustration purposes.}. Channel execution 1 does not examine t100 because the tuple is not in storage yet, and channel execution 2 will not examine t100 either, because the tuple has an ingested\_timetsamp that is smaller than 10 (i.e., too old). Thus, tuple t100 is missed.
    
    \begin{figure}[h]
        \centering
        \includegraphics[width=0.90\textwidth]{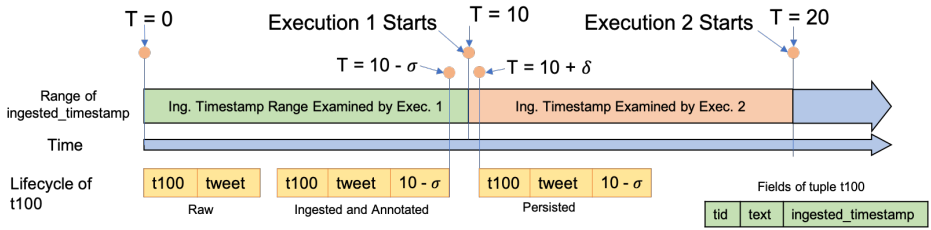}
        \caption{Missing tuple due to early timestamping}
        \label{fig:early_timestamping}
    \end{figure}

    \item \textit{Inappropriate Data Access:} 
    We have attached an explicit timestamp (``ingested\_timestamp'') attribute to mark the ingestion time of incoming tweets.
    This attribute then exists as part of the user data, and other users of the BAD system can access it. This raises the potential risk that other users may accidentally modify this attribute and cause data channels to fail. Additionally, this auxiliary information may cause confusion for non-channel users such as data analysts.
    \item \textit{Complex Approximation Query:} 
    In order to approximate continuous semantics, we have chosen the same time period in the temporal predicate and the channel execution period, as shown in Figure~\ref{ddl:new_nearby_hateful_tweets}. Such a correspondence needs to be managed manually and carefully by developers. When channel queries become more complex and involve multiple incoming data sources, constructing a proper approximation query can be challenging. One would have to add proper temporal predicates for each of the data sources, and when there are joins between these data sources, which portion of the collected data from one data source should be joined the other one needs to be carefully specified with temporal predicates. These temporal predicates would increase the query complexity and make such queries very difficult to write.
\end{itemize}

The above challenges of using BAD-RQ to approximate continuous semantics introduce risks of missing data and cause difficulties for developers in
creating continuous BAD applications. In order to properly support continuous BAD, we introduce a new BAD service - BAD-CQ - with native support for continuous query semantics.

\subsection{BAD-CQ}
\label{sec:bad-cq}
In this section, we first introduce the new building blocks needed for providing continuous semantics in BAD-CQ, and then we show how to utilize them to create continuous data channels for continuous BAD.

\subsubsection{Active Datasets}
\label{sec:active_dataset}
\begin{figure}[h]
    \centering
    \includegraphics[width=0.60\textwidth]{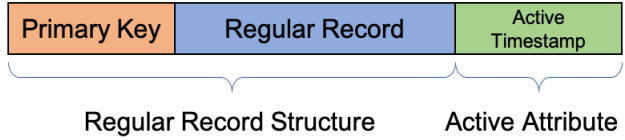}
    \caption{Storage format of an active record}
    \label{fig:active_record}
\end{figure}

As we have discussed in Section~\ref{sec:bad_timestmaps}, BAD persists all data to support retrospective analysis. To help data channels differentiate new data from old, we need to timestamp incoming data and use timestamps for proper continuous channel evaluation. To avoid the previously mentioned drawbacks of adding an ingestion timestamp to user data, we introduce a new type of datasets - \textit{Active Datasets} - in BAD-CQ. Unlike regular datasets in AsterixDB, a record (active record) stored in an active dataset contains not only user data but also a ``hidden'' active attribute: ``\_active\_timestamp''. This helps BAD-CQ to evaluate continuous channel queries. This attribute is stored alongside users' data but separated from the regular record content. It is ``invisible'' to users and can only be accessed using active functions (to be discussed soon). The storage layout of an active record is shown in Figure~\ref{fig:active_record}.

As the BAD system runs in a distributed environment, which clock to use to assign active timestamps needs careful consideration. One might first consider using a single clock to assign all active timestamps. This would be convenient because then all active timestamps would be directly comparable, and we would only need to figure out one active timestamp range to identify all the new data. However, having a master clock would require either routing all data to a single node, which would create a bottleneck in the system, or synchronizing clocks on multiple nodes, which can be very 
challenging in a distributed environment.
In BAD-CQ, we instead use the local clock on each node to assign active timestamps to the active records stored on it for scalability. Active timestamps are assigned in the storage engine, after the locking phase. This makes sure that incoming records will become visible to running queries as soon as they are timestamped.
Although the \textbf{new data} on each node may now have
a different active timestamp range, we can introduce an active timestamp management mechanism with additional query optimization rules to make sure that channel queries are evaluated correctly. We will further discuss this in Section~\ref{sec:bad_cq_syntax_and_opt}.

\begin{figure}[h]
    \centering
    \includegraphics[width=0.65\textwidth]{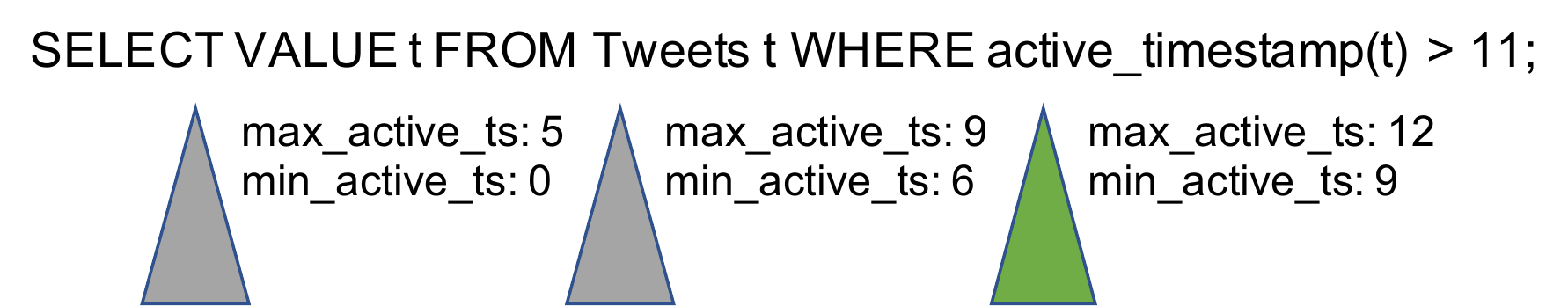}
    \caption{Access active datasets with filters}
    \label{fig:active_ts}
\end{figure}

Considering that active timestamps often need to be compared in channel queries, we can optimize these comparisons to improve channel performance.
One might consider creating a secondary index on active timestamps, but this would take additional disk space and incur additional access overhead when the selectivity is high~\cite{luoc_lsm}. 
As the active timestamps of an active dataset grow monotonically, we can instead utilize the filter feature in the AsterixDB storage engine to avoid accessing irrelevant data~\cite{asterixdb_filter}. The BAD storage engine uses Log-Structured Merge (LSM) Trees as its storage structure~\cite{asterixdb_storage}; they perform batch updates into components to avoid the cost of random writes and then read them sequentially for data access. One can designate a filter attribute when creating a dataset, and every LSM component of this dataset is then decorated with the maximum and minimum attribute values of its stored records. When a query containing a filter attribute comparison comes, it can quickly skip irrelevant components by examining their maximum and minimum filter values. For active datasets, we use the active timestamp as the filter attribute to accelerate channel queries, as shown in Figure~\ref{fig:active_ts}. The \textit{active\_timestamp(t)} function reveals the active timestamp of the tuple t stored in the active dataset \textit{Tweets}, as will further be discussed in Section~\ref{sec:bad_cq_syntax_and_opt}.

The syntax for creating active datasets is straightforward. An active dataset can be created with a regular data type, and the active attribute and filter are automatically configured behind-the-scene. One can create two active datasets \textit{Tweets} and \textit{OfficerLocations} using the statements in Figure~\ref{ddl:create_active_datasets}. Active datasets can also be accessed in regular queries just like non-active datasets. There is an extra overhead when reading active datasets due to the additional space for storing active timestamps. We will see from later experiments that this overhead is relatively small. When not used in query evaluation, active timestamps are projected out from the active records as early as possible to avoid potential transmission overhead.

\begin{figure}[h]
\footnotesize
\centering
\begin{lstlisting}[
           language=SQL,
           basicstyle=\ttfamily,
           showstringspaces=false,
           commentstyle=\color{gray}
        ]
  CREATE ACTIVE DATASET Tweets(Tweet) PRIMARY KEY oid;
  CREATE ACTIVE DATASET OfficerLocations(OfficerLocation) PRIMARY KEY oid;
\end{lstlisting}
\caption{Datatype and dataset definition for officer location updates}
\label{ddl:create_active_datasets}
\end{figure}

\subsubsection{Active Timestamp Management}
\label{sec:active_timestamp_management}
With active datasets, we now need to ``teach'' channels to utilize the active timestamps to recognize \textbf{new} data and to guarantee continuous semantics. The basic idea is straightforward: keep track of the channel execution times and compare them with active timestamps to find the new data. As mentioned in Section~\ref{sec:active_dataset}, 
each node uses the local time
to assign active timestamps, so we also need to use local time for tracking channel execution times and make sure they are properly compared with active timestamps. We create a local active timestamp manager on each node to keep track of the \textit{previous channel execution time} and the \textit{current channel execution time} under the local clock. When a channel executes on a node, 
these two timestamps are used to determine which portion of the stored data should be considered for this execution.

\begin{figure}[h]
    \centering
    \includegraphics[width=\textwidth]{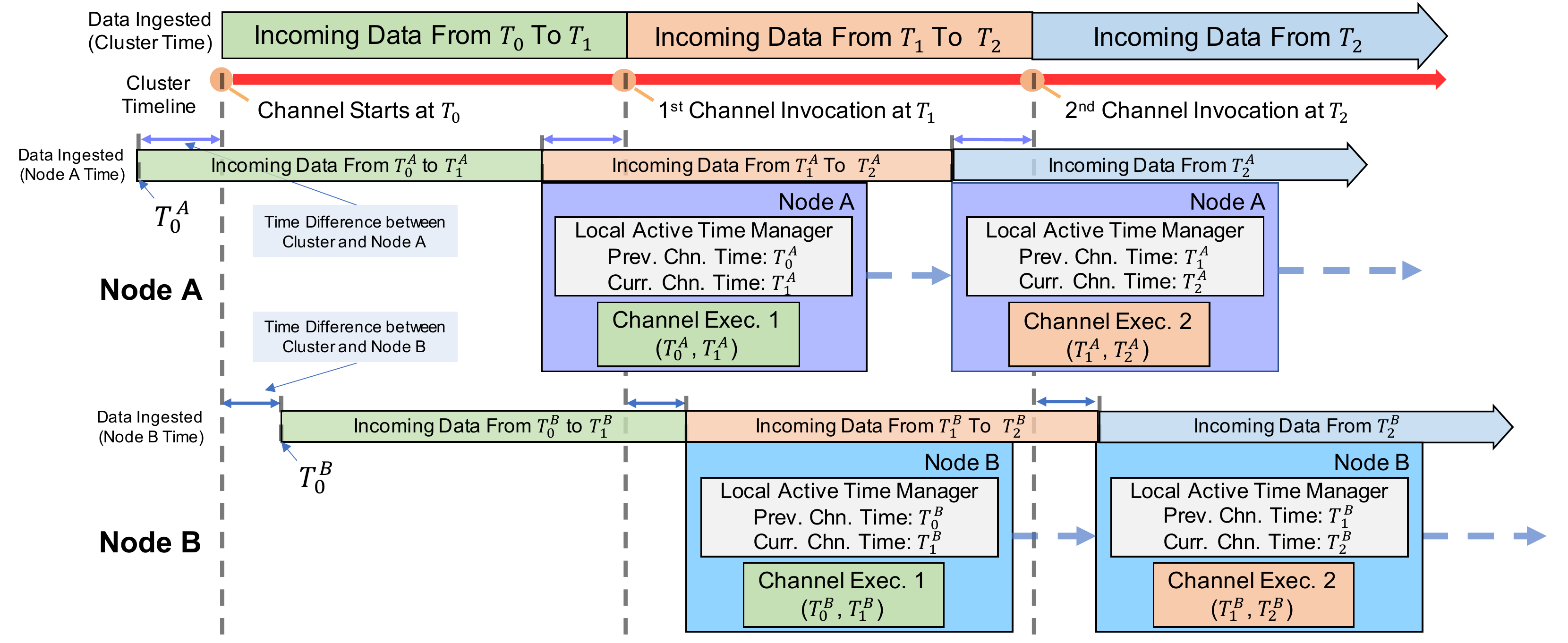}
    \caption{An illustration of active timestamp management}
    \label{fig:active_timestamp_management}
\end{figure}

To demonstrate how multiple local active timestamp managers can work to offer continuous semantics, we consider the channel defined in Section~\ref{sec:approx_c} that looks for \textbf{new} nearby tweets for in-field officers. We show an illustrative channel execution example in Figure~\ref{fig:active_timestamp_management}. In this example, we use the cluster controller (CC) time as the (conceptual) cluster time. Since not all nodes are synchronized on time, current timestamps on different nodes can be different. In this case, when CC starts the first channel execution at time $T_0$, Node A marks the channel start time under its local time as $T_0^A$, which is \textit{``logically before''} $T_0$, and Node B marks the channel start time under its local time as $T_0^B$, which is \textit{``logically after''} $T_0$. 
When the CC invokes the first channel execution at $T_1$, every node examines the tweets ingested and persisted from the previous channel execution time to the current channel execution time. From Node A's perspective, all tweets ingested from $T_0^A$ to $T_1^A$ are examined. From Node B's perspective, tweets ingested from $T_0^B$ to $T_1^B$ are examined. Although $T_1$, $T_1^A$, and $T_1^B$ are different, from the CC's (and subscribers') perspective, only nearby hateful tweets from $T_0$ to $T_1$ are reported to subscribers. This guarantees the continuous semantics for this channel. 
The channel's previous channel execution and current channel execution time are each progressed with each channel execution. They are updated instantly when a channel execution job first accesses an active dataset used for the channel. This makes sure that all incoming tweets that were persisted before the current channel execution can all be safely examined in the current channel execution.

The active timestamp manager enables BAD-CQ to provide continuous semantics in a distributed environment without time synchronization. The monotonically increased active timestamps on each node in fact act like sequence numbers. The local active time manager marks the range of sequence numbers for each channel execution (as its previous and current channel execution time) and allows it to find the new data.

\subsection{BAD-CQ Syntax and Optimization}
\label{sec:bad_cq_syntax_and_opt}
Active datasets and active timestamp management allows BAD-CQ to provide continuous semantics.
In order to enable users to use active timestamps and channel execution times for constructing channel queries, we introduce several active functions in this section. 
Each active function takes a parameter that refers to tuples from active datasets. 
Applying active functions on normal datasets will cause a query compilation exception. In order to describe the functionalities of active functions, we use a tuple \textit{t} from the active dataset \textit{Tweets} as an example. The active functions are as follows:

\begin{itemize}
    \item \textit{\textbf{active\_timestamp(t)}} reveals the active timestamp of the tuple \textit{t}.
    \item \textit{\textbf{previous\_channel\_time(t)}} returns the previous channel execution time on the node where the tuple \textit{t} is persisted, as defined in Section~\ref{sec:active_timestamp_management}. Note that every node has its own (local) channel time for a channel, and dataset \textit{Tweets}'s tuples could be persisted on multiple nodes, so this function is evaluated locally on each node at run time, and tuples from Tweets used in the channel could have different previous channel times. 
    \item \textit{\textbf{current\_channel\_time(t)}} returns the current channel execution time of the tuple \textit{t}, as defined in Section~\ref{sec:active_timestamp_management}. Similar to \textit{previous\_channel\_time}, \textit{current\_channel\_time} is also computed locally at run time, and tuples from \textit{Tweets} could have different current channel time.
    \item \textit{\textbf{is\_new(t)}} returns a boolean value indicating whether tuple \textit{t} is new to the current channel execution. The return value of \textit{is\_new(t)} is equivalent to the following expression:\newline
    \small{\textit{previous\_channel\_time(t) $<$ active\_timestamp(t) AND active\_timestamp(t) $<$ current\_channel\_time(t)}}.
\end{itemize}

With active functions, a developer can conveniently construct continuous channels with continuous semantics. Here we show an example for the use case described in Section~\ref{sec:approx_c}, where subscribers would like to receive new tweets near in-field officers. We use a different user model in BAD-CQ. Data channel definition in BAD-CQ is not based on UDFs, since active functions are not meaningful outside. Executing \textit{previous\_channel\_time} and \textit{current\_channel\_time} functions in regular queries return 0 and current cluster time respectively. 
Using BAD-CQ's active functions, a developer can create a continuous channel for the new nearby hateful tweets using the statement shown in Figure~\ref{ddl:new_nearby_hateful_tweets_cq}.

\begin{figure}[h]
    \centering
    \includegraphics[width=0.65\textwidth]{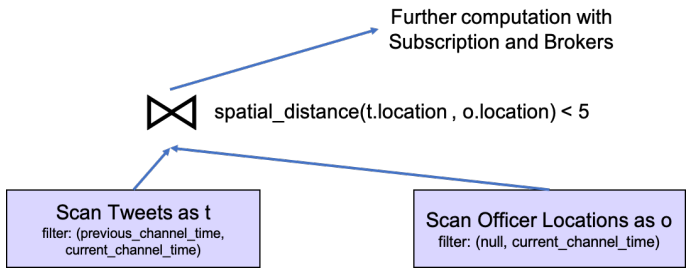}
    \caption{Query plan for new nearby hateful tweet channel}
    \label{fig:new_nearby_hateful_tweets_plan}
\end{figure}

In order to assist channel evaluation with active functions and to improve channel performance, we introduce two new query optimization rules into BAD-CQ. First, when compiling a continuous channel query, we push the  \textit{current\_channel\_time} function into the leaf node of the query plan - the data scan operator of an active dataset - as the filter's maximum value. This is because when an active dataset is accessed in a channel execution, only data before the current channel execution time is relevant. We use this to quickly skip data coming after the current execution starts. 
Second, we push the \textit{previous\_channel\_time} function down towards the leaf of the query as much as possible, and we use it as the filter's minimum value for active datasets when applicable. Whether this function can be pushed into the data scan operator depends on the specific channel query. For the channel query defined in  Figure~\ref{ddl:new_nearby_hateful_tweets_cq}, we can indeed push \textit{previous\_channel\_time(t)} into the \textit{Tweets} scan operator and use it as the minimum filter, as shown in Figure~\ref{fig:new_nearby_hateful_tweets_plan}.~\footnote{
In this channel, we only need officers' latest location, so there is no lower bound on active timestamps of \textit{OfficerLocations}. We will introduce another example in Section~\ref{sec:bad_cq_semantics} which requires recent location updates and utilizes the minimum filter on \textit{OfficerLocations}.}

\begin{figure}[h]
\footnotesize
\begin{lstlisting}[
           language=SQL,
           basicstyle=\ttfamily,
           showstringspaces=false,
           commentstyle=\color{gray}
        ]
CREATE CONTINUOUS CHANNEL CQNewNearbyHatefulTweets(oid) PERIOD duration("PT10S") {
  SELECT t
  FROM OfficerLocations o, Tweets t
  WHERE spatial_distance(t.location, o.location) < 5 
    AND o.oid = oid AND t.hateful_flag = true AND is_new(t)
};
\end{lstlisting}
\caption{A continuous channel for new nearby hateful tweets}
\label{ddl:new_nearby_hateful_tweets_cq}
\end{figure}

When the \textit{previous\_channel\_time} function cannot be pushed all the way down into a data scan operator, we need to attach its node-dependent value (i.e., the previous channel execution time on a node) to the active records read from this node. In this case, the comparison between active timestamps and the \textit{previous\_channel\_time} function is rewritten into a comparison between active timestamps and this attached previous channel execution time value. This makes sure that even if active records are shuffled around in the cluster, the comparison between their active timestamps and previous channel time will be evaluated correctly. To explain how 
the second rule works in this scenario, we introduce another continuous use case, where \textit{``in-field officers (as subscribers) would like to receive nearby hateful tweets he/she has not seen before''}. In this case, we not only need to consider a new tweet posted near an in-field officer, but also tweets that were not nearby but that become nearby due to the movement of in-field officers. We can create a continuous channel for this use case as shown in Figure~\ref{ddl:unseen_nearby_hateful_tweets_cq}.

\begin{figure}
\scriptsize
\begin{lstlisting}[
           language=SQL,
           basicstyle=\ttfamily,
           showstringspaces=false,
           commentstyle=\color{gray}
        ]
  CREATE CONTINUOUS CHANNEL UnseenNearbyHatefulTweets(oid) PERIOD duration("PT10S") {
    SELECT t
    FROM OfficerLocations o, Tweets t
    WHERE spatial_distance(t.location, o.location) < 5 AND o.oid = oid 
      AND t.hateful_flag = true 
      AND (is_new(o) OR is_new(t))
  };
\end{lstlisting}
\caption{A continuous channel for unseen nearby hateful tweets}
\label{ddl:unseen_nearby_hateful_tweets_cq}
\end{figure}

\begin{figure}
    \centering
    \includegraphics[width=0.90\textwidth]{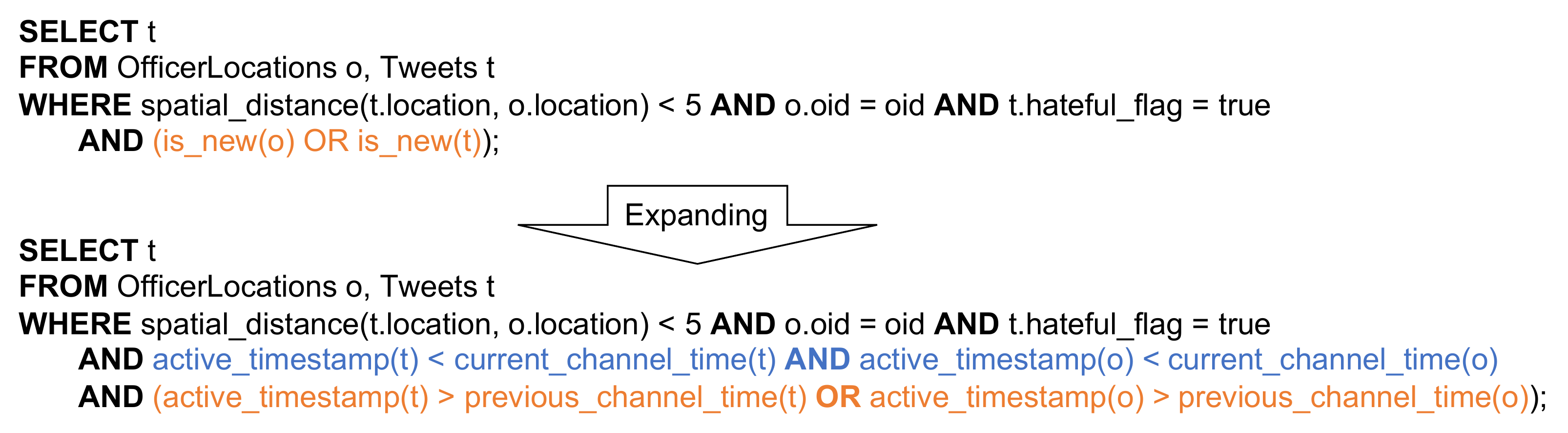}
    \caption{Expanding a continuous channel query with active functions}
    \label{fig:active_func_unfolding}
\end{figure}

\begin{figure}[h]
    \centering
    \includegraphics[width=0.65\textwidth]{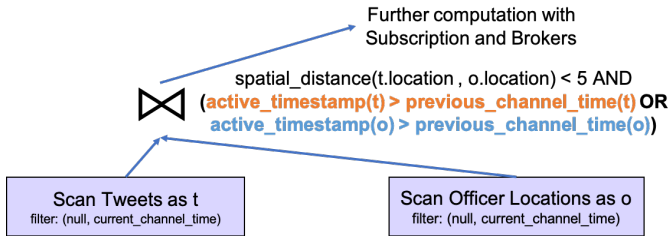}
    \caption{Query plan for unseen nearby hateful tweet channel}
    \label{fig:unseen_nearby_hateful_tweets_plan}
\end{figure}

In this continuous channel query, the active functions \textit{is\_new(o)} and \textit{is\_new(t)} are expanded to the corresponding query predicates based on active timestamps, previous channel execution time, and current channel execution time as shown in Figure~\ref{fig:active_func_unfolding}. Following the first optimization rule, the current time timestamp of both \textit{Tweets} and \textit{OfficerLocations} are pushed into the corresponding data scan operators. However, the previous channel execution time cannot be pushed thoroughly, because the disjunctive predicate ``\textit{active\_timestamp(t) $>$ previous\_channel\_time(t) OR active\_timestamp(o) $>$ previous\_channel\_time(o)}'' also needs data from before the previous channel execution time from both datasets. Following the second optimization rule, this continuous channel query can be compiled into the plan shown in Figure~\ref{fig:unseen_nearby_hateful_tweets_plan}. The disjunctive predicate is evaluated in the join operation that is computed across all nodes, and data is shuffled around in this process~\footnote{Depending on the workload, the execution plan for the channel query can choose either to broadcast \textit{Tweets} or \textit{OfficerLocations}.}.
Notice now that since the previous channel execution time is attached to active records, we can compare the active timestamp with the channel execution time under the same local clock, even if records are shipped to another node.

Different from the implicit query rewriting in Tapestry~\cite{tapestry} and the delta files in NiagaraCQ~\cite{niagara_cq}, 
BAD-CQ allows developers to construct queries using active functions that are best suited for their use cases, and it takes advantage of the storage engine for accelerating channel queries without having to introduce additional data structures.
Developers can write a query using the \textit{is\_new} function and let the query compiler rewrite it into an incremental query, or they can use the \textit{active\_timestamp} function to expose the active timestamps and directly compare them with channel times or other times. The BAD-CQ user model uses datasets to hold the collected incoming data and other existing data. This provides developers with a unified query model and lets them to reuse all dataset processing operations when defining channels. The principles underlying the BAD-CQ approach are general - i.e., other database systems supporting declarative queries could also be adapted to provide continuous semantics like BAD-CQ.

\subsection{BAD-CQ Semantics}
\label{sec:bad_cq_semantics}
To better
understand the query semantics provided in BAD-CQ, we dive into the details of several continuous BAD use cases in this section. We focus on the scenario where in-field officers would like to get nearby hateful tweets with different preferences, and we use data samples to show how BAD-CQ produces notifications for different channels.

\subsubsection{New Nearby Hateful Tweets}
\label{sec:bad_cq_semantics_1}
We first look at the example from Section~\ref{sec:approx_c}, where in-field officers would like to receive new nearby hateful tweets. The channel is defined in Figure~\ref{ddl:new_nearby_hateful_tweets_cq}. 
We use the \textit{is\_new} function to look for \textbf{new} tweets that have not been sent to subscribers, and we use the officers' latest locations to look for nearby tweets.

\begin{figure}[h]
    \centering
    \includegraphics[width=0.85\textwidth]{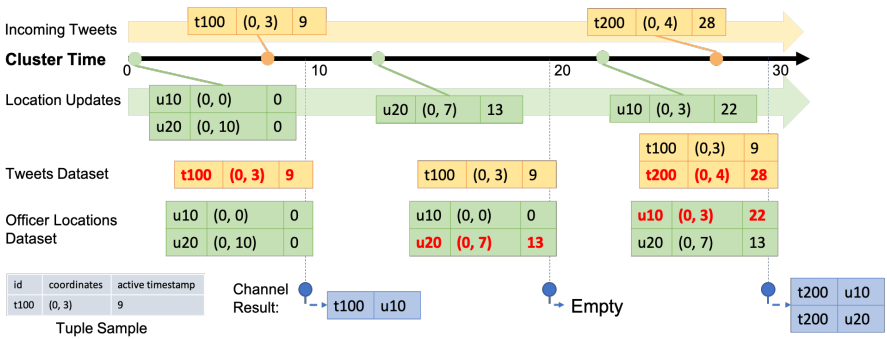}
    \caption{Officer u10 subscribing to CQNewNearbyHatefulTweets(u10) and officer u20 subscribing to CQNewNearbyHatefulTweets(u20) }
    \label{fig:recent_tweets_example}
\end{figure}

In Figure~\ref{fig:recent_tweets_example}, we show a channel execution example with several sample data records. In order to focus on the channel execution process, irrelevant attributes of tweets and officer location updates are not shown in the figure. The channel starts at time 0, and in-field officers u10 and u20 have initial location at time 0 of (0, 0) and (0, 10), respectively. At 9s, the first tweet t100 arrives and its location is (0, 3). When the channel first executes at 10s, only tweet t100 is near in-field officer u10, so the channel produces one notification for u10. After that, u20 updates his/her location to (0, 7) at 13s. When the channel executes at 20s, as there is no new tweet after the previous channel execution, no notification is produced. Later, u10 updates his/her location to (0, 3) at 22s, and a new tweet t200 located at (0, 4) arrives at 28s. When the channel executes at 30s, both u10 and u20 have t200 nearby, so the channel produces two notifications for each of the corresponding officers.

\subsubsection{Unseen Nearby Hateful Tweets}
In the previous use case, in-field officers receive a hateful tweet only if the tweet is temporally new. In another use case, officers may also be interested in older nearby hateful tweets that they have not seen before (which could contain useful information). The channel definition for this use case is shown in Figure~\ref{ddl:unseen_nearby_hateful_tweets_cq}.

\begin{figure}[h]
    \centering
    \includegraphics[width=0.85\textwidth]{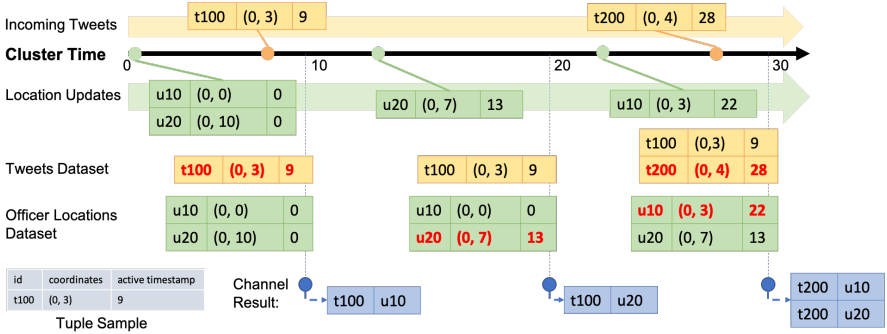}
    \caption{Officer u10 subscribing to UnseenNearbyHatefulTweets(u10) and officer u20 subscribing to UnseenNearbyHatefulTweets(u20)}
    \label{fig:unseen_tweets_example}
\end{figure}

We use the same data sample in Section~\ref{sec:bad_cq_semantics_1} to explain how this channel works. As shown in Figure~\ref{fig:unseen_tweets_example}, the channel acts the same way as the previous one and produces one notification for u10 in the first channel execution. In the second channel execution, the location update of u20 from (0, 10) to (0, 7) makes t100 become nearby to u20, so the channel produces one notification for u20 at 20s to notify this officer about this previously unseen tweet. The third channel execution starts at 30s and produces two notifications for u10 and u20, respectively, as both in-field officers have not seen this new tweet.

\subsubsection{New Nearby Hateful Tweets for Active Officers}
\label{sec:new_nearby_hateful_tweets_for_active_officers}
In the previous use cases, even if an officer is not updating his/her location constantly (e.g., in order to reduce power/data plan consumption), the channel can still be producing notifications for them based on their last known location. When the officer reconnects, the broker sub-system can pull notifications that were produced ``offline'' from the BAD storage engine and send them out. 
If we want to produce notifications only to ``active'' in-field officers (who are their updating the locations to the system regularly), one can create the continuous channel defined in Figure~\ref{ddl:new_nearby_hateful_active_user}. Different from the channel defined in Figure~\ref{ddl:new_nearby_hateful_tweets_cq}, we now only look for new hateful tweets for officers who have recently updated their locations instead of all officers. Those who are not updating their locations ``actively'' will not receive nearby hateful tweets while they are inactive.

\begin{figure}[h]
\scriptsize
\begin{lstlisting}[
           language=SQL,
           basicstyle=\ttfamily,
           showstringspaces=false,
           commentstyle=\color{gray}
        ]
  CREATE CONTINUOUS CHANNEL NewNearbyHatefulTweetsForActiveOfficers(oid) 
   PERIOD duration("PT10S") {
    SELECT t
    FROM OfficerLocations o, Tweets t
    WHERE spatial_distance(t.location, o.location) < 5 
     AND o.oid = oid AND t.hateful_flag = true AND is_new(t) AND is_new(o)
  };
\end{lstlisting}
\caption{A continuous channel for new nearby hateful tweets}
\label{ddl:new_nearby_hateful_active_user}
\end{figure}

\begin{figure}[h]
    \centering
    \includegraphics[width=0.85\textwidth]{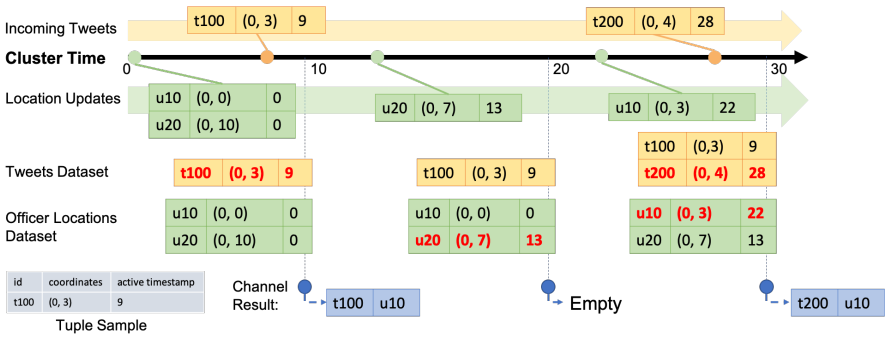}
    \caption{Officer u10 subscribing to NewNearbyHatefulTweetsForActiveOfficers(u10) and officer u20 subscribing to NewNearbyHatefulTweetsForActiveOfficers(u20)}
    \label{fig:recent_tweets_active}
\end{figure}

Following our data sample used in previous use cases, the execution process of this channel is shown in Figure~\ref{fig:recent_tweets_active}. Similarly, the first channel execution produces one notification based on u10 about t100. In the second channel execution, no notification is generated since there is no new incoming tweet. In the third channel execution, we produce one notification about the new tweet t200 for u10 who has recently updated his/her location.
Although t200 is also near u20, we do not produce a notification for him/her since u20 is not ``active''. As we can see from these sample use cases, active functions offer the flexibility and expressiveness of working with both the \textit{new} and \textit{historical} data. Developers can use active functions to conveniently construct a wide range of suitable queries for their BAD applications.

\section{GOOD: A Not BAD Approach}
\label{sec:not_bad}
In order to fully support BAD applications without the BAD system, one would have to glue multiple existing Big Data systems together. In this section, we discuss a Not-BAD approach, which we call GOOD
- \textbf{G}luing \textbf{O}odles \textbf{O}f \textbf{D}ata platforms -
approach to approximate the BAD system. We introduce a GOOD system that consists of several Big Data systems, illustrate how to configure it for creating BAD services, and compare it with the BAD system.

\subsection{The GOOD Architecture}
Following our discussion in Section~\ref{sec:a_bad_world}, a GOOD system also needs to serve all three types of BAD users: \textbf{Subscribers} who want to customize data and receive constant updates, \textbf{Developers} who create BAD applications to serve subscribers, and \textbf{Analysts} who analyze data using declarative queries. Such a system should provide the following features: 

\begin{itemize}
    \item Efficient data ingestion for rapid incoming data.
    \item Data customization based on a large volume of subscriptions.
    \item Data analytics with a declarative language.
    \item Persistent storage for incoming data and other relevant information.
    \item Customized data delivery to a large number of subscribers.
\end{itemize}

An existing Big Data system alone can only fulfill a portion of the BAD requirements. For example, Apache Spark Structured Streaming offers on-the-fly data processing but lacks persistent storage. Amazon's Simple Notification Service (SNS) supports cloud-based pub/sub, but the expressiveness of subscriptions is limited to the content of publications.
A user wanting to build BAD applications would thus have to glue multiple systems together. We can break down a proposed GOOD system architecture into different components and categorize existing Big Data systems with respect to this GOOD architecture, as shown in Figure~\ref{fig:good_taxonomy}. We describe the functionality of each component as follows:

\begin{figure}[h]
    \centering
    \includegraphics[width=\textwidth]{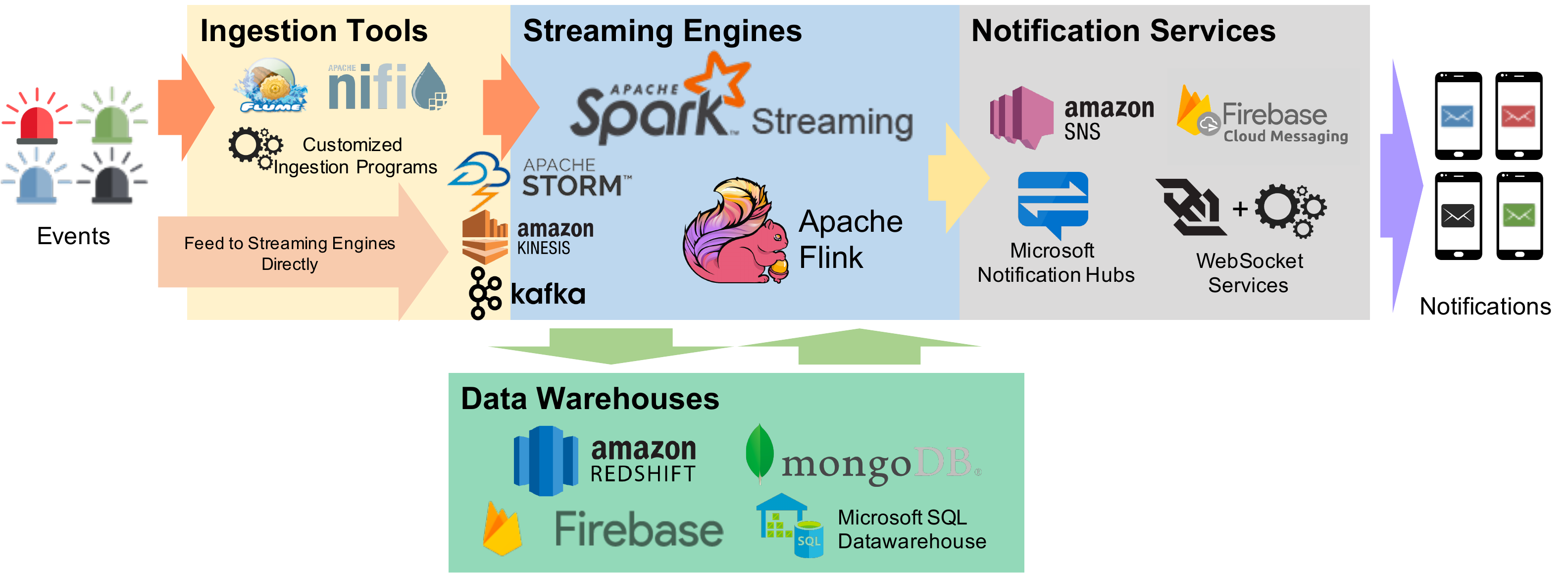}
    \caption{GOOD Architecture}
    \label{fig:good_taxonomy}
\end{figure}

\begin{itemize}
    \item \textbf{Ingestion tools} collect data from external data sources and help distribute the data to downstream components. In some cases, users could implement their own ingestion programs to handle specific ingestion protocols. With the growth of stream processing, many ingestion tools now also support on-the-fly data processing (with various limitations). This blurs the boundary between ingestion tools and streaming engines. Here we still consider them as
    different GOOD components to emphasize their functional differences.
    \item \textbf{Streaming engines} today come in two different flavors. One (e.g., Apache Storm, Apache Kafka) focuses on efficient and reliable data distribution and allows users to hang data processing units onto the pipeline. The other (e.g., Apache Flink, Apache Spark Structured Streaming) focuses on enabling real-time data analytics as if working with non-streaming data. Users could glue multiple streaming engines together to benefit from both flavors (such as gluing Kafka with Spark Structured Streaming). GOOD can use streaming engines to combine incoming data, subscription information, and other relevant data to produce customized notifications.
    \item \textbf{Data warehouses} provide data persistence and support data analytics. We use data warehouses as the storage engine of the GOOD system. Incoming data is persisted in data warehouses for retrospective analysis. Subscriptions and other relevant data used for producing customized data are also persisted in data warehouses and loaded into streaming engines for processing. We do not replicate data in both streaming engines and data warehouses to avoid data inconsistency and the constant migration of updates between them.
    \item \textbf{Notification services} deliver customized data produced by streaming engines to interested subscribers. Users could choose cloud-based services such as Amazon SNS or Firebase Cloud Messaging to send notifications to subscribers via SMS or Email, or they also could build their own notification services based on WebSocket.
\end{itemize}

Every component of the GOOD system must be horizontally scalable to ensure that it can support a large number of subscribers, just like the BAD system. Even with this scalable architecture, it would be impractical for the GOOD system to compute/customize an incoming data item for every subscriber independently, especially when the incoming data arrives rapidly. In order to best approach the BAD system's scalability requirement, we also adopt the data channel model in our GOOD system architecture by grouping similar subscriptions into a data channel and evaluating them together. 
Next, we will consider a sample GOOD system to explain how it can receive, customize, and deliver data.

\subsection{A GOOD System}
\label{sec:a_good_system}
The GOOD architecture offers a way to approximate the BAD system by gluing multiple existing Big Data systems together. 
One could choose various combinations among the options in Figure~\ref{fig:good_taxonomy} for creating a GOOD system. In order to compare the GOOD system with the BAD system toe-to-toe, we have constructed a sample GOOD system using several component systems that have been widely used in practice, as shown in Figure~\ref{fig:a_good_system}. 
We choose Apache Kafka for data ingestion and use Spark Structured Streaming for data processing, as suggested in the Spark Structured Streaming documentation~\cite{spark_streaming_doc}. Although Kafka also supports several data processing operations via Kafka Streams~\cite{link:kafka_streams}, we choose Spark Structured Streaming for its richer query semantics, which is closer to the BAD system's offering. 
Further, we use MongoDB for persistence and analytics and AmazonSNS for notification delivery.
Each component of the GOOD system can be described and configured as follows:

\begin{figure}[h]
    \centering
    \includegraphics[width=0.85\textwidth]{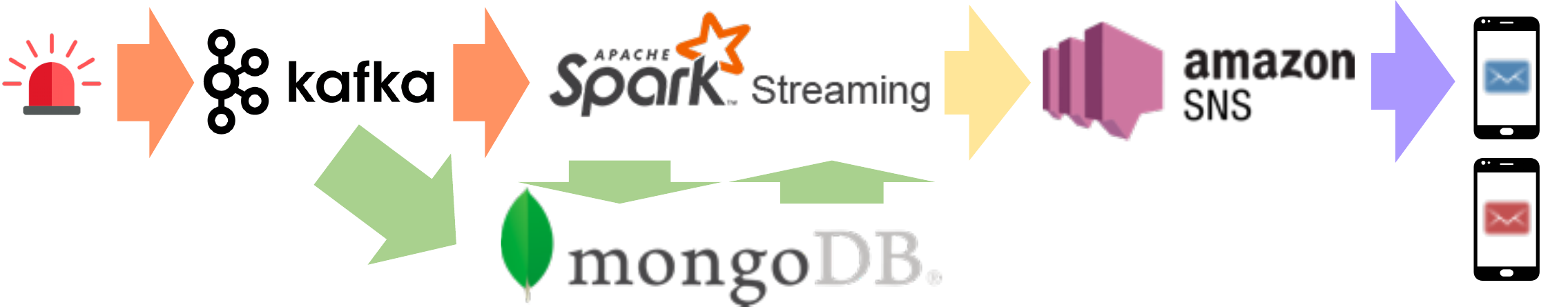}
    \caption{A concrete GOOD system}
    \label{fig:a_good_system}
\end{figure}

\begin{itemize}
    \item \textbf{Apache Kafka} is a distributed streaming platform that allows applications to publish and subscribe to data streams reliably. We connect external data sources to Kafka using producer APIs. For each data source, we can create a topic in Kafka to allow downstream consumers (Spark Structured Streaming and MongoDB) to access the incoming data.
    \item \textbf{MongoDB} is a document-based distributed database. We connect MongoDB to Kafka as a consumer via the mongodb-kafka connector~\cite{link:kafka_mongo} provided by MongoDB. Incoming data records from a Kafka topic (i.e., an external data source) are persisted in a corresponding MongoDB collection as JSON-like documents for retrospective analysis. Besides incoming data, subscriptions specifying subscribers' interest and other relevant information used for data customization and data analytics are also stored in MongoDB.
    \item \textbf{Apache Spark Structured Streaming} is a scalable stream processing engine built on top of the Spark SQL engine. It supports Dataframe/Dataset APIs for users to express streaming computations the same way one would express a batch computation on static data. We connect Spark Structured Streaming to Kafka as a consumer through the spark-streaming-kafka connector~\cite{link:spark_kafka} provided by Spark. Incoming data from a Kafka topic is mapped into a data stream in Spark Structured Streaming. One can implement a data channel as a Spark application that runs continuously for producing customized data. Relevant information and subscriptions stored in MongoDB can be loaded into Structured Streaming as DataFrames through a mongodb-spark connector~\cite{link:mongo_spark} provided by MongoDB.
    \item \textbf{Amazon SNS} is a notification service provided in Amazon Web Services for delivering messages to subscribed endpoints or clients. It allows users to create Amazon SNS topics and publish notifications through APIs. Other systems and end-users can subscribe to these topics and receive published data. Amazon SNS provides filter policies in subscriptions to allow subscribers to filter notifications by their content. We can use the filter policy to send notifications to certain channel subscribers by using their subscription IDs as the filter value. We map a data channel to an Amazon SNS topic, and whoever subscribes to this data channel also becomes a subscriber to the Amazon SNS topic with its subscription ID as the filter attribute. Customized data generated by the Spark channel application is published to the Amazon SNS topic with subscription IDs attached.
\end{itemize}

Due to its glued nature, the GOOD system needs ``cooperation'' between different components to provide BAD services. Taking the new nearby hateful tweet example described in Section~\ref{sec:continuous_bad} (the equivalent BAD channel defined in Figure~\ref{ddl:new_nearby_hateful_tweets_cq}), one would have to complete the following steps for providing the channel service in the GOOD system:

\begin{itemize}
    \item Configure and deploy Apache Kafka to the cluster.
    Create adaptor programs as Kafka producers that publish data into Kafka topics for tweets and for officer location updates separately.
    \item Configure and deploy MongoDB to the cluster. Create collections for tweets, location updates, and subscriptions, and make sure all collections are sharded across the cluster.
    \item Create and configure an Amazon SNS topic on Amazon Web Services for sending notifications.
    \item Configure and deploy Apache Spark to the cluster. Create a Spark application as a data channel and connect it to Kafka, MongoDB, and Amazon SNS separately. Implement data customization by joining tweets, officer locations, and subscriptions using stream processing operations.
    \item Deploy the channel application onto the Spark cluster and make sure all services are running and connected.
    \item For each newly subscribed subscriber, we add the subscription information into MongoDB for data customization, and we also create a corresponding Amazon SNS subscription with the subscription ID as the filter attribute.
\end{itemize}

Compared with the BAD system, the GOOD system requires a significant amount of effort from developers to configure, orchestrate, and manage different components for providing BAD services. Besides the administration complexity, due to the limitation of the components in the GOOD system, not all of the query semantics provided by the BAD system can be conveniently supported by the GOOD system.

\subsection{GOOD vs. BAD}
\label{sec:good_vs_bad}
As we have mentioned, streaming engines have to age historical data out to restrain their resource usage. This limits the query semantics that can be supported by the GOOD system. Consider the new nearby hateful tweets channel defined in Figure~\ref{ddl:new_nearby_hateful_tweets_cq}, where we send new nearby hateful tweets to in-field officers based on their last known location by utilizing an UPSERT feed.
That channel can produce notifications for a temporarily ``offline'' officer and later send these ``missed'' notifications to him/her when the officer reconnects, as discussed in Section~\ref{sec:bad_cq_semantics}.

In the GOOD system, if an officer has not sent location updates for a some time, his/her location information would be aged out by the streaming engine. Due to this limitation, a GOOD user can only look for location updates back to a limited time for a new incoming tweet. To better approximate the BAD channel, one could consider persisting all historical location updates in MongoDB and pulling the latest locations into Spark Structured Streaming in each channel execution. However, this would lose the timeliness of streaming data and introduce additional data access overhead.

\begin{figure}[h]
    \centering
    \includegraphics[width=0.85\textwidth]{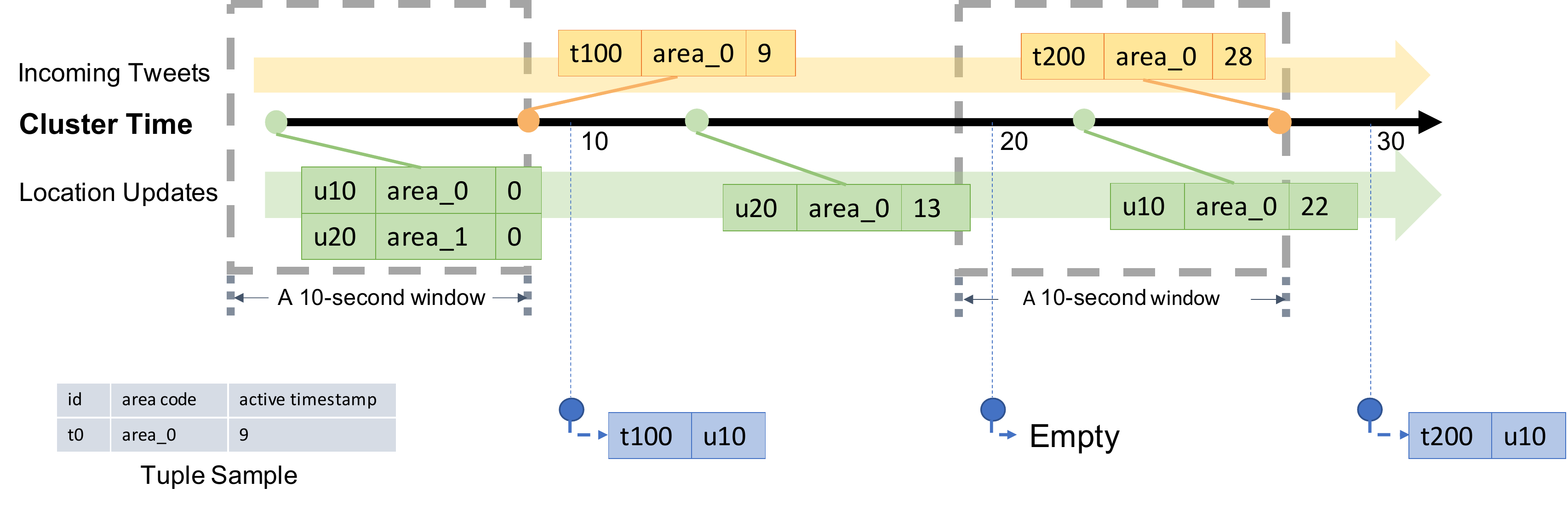}
    \caption{A GOOD example of sending hateful tweets to officers}
    \label{fig:recent_tweets_good}
\end{figure}

To illustrate the query semantics of the GOOD system and compare that with BAD, we show an alternative new nearby hateful tweets use case. As Spark Structured Streaming does not support spatial joins on data streams, we use ``area\_code'' to represent tweets' and officers' locations. We consider a tweet to be nearby to an officer if it is posted from the same area code as the officer. In this modified use case, we send a new hateful tweet to the nearby in-field officers who have recently (within 10 seconds) updated their locations. This use case will also be used in the later performance comparison between BAD and GOOD. An illustrative example of the modified channel execution using the data sample in Section~\ref{sec:bad_cq_semantics} is shown in Figure~\ref{fig:recent_tweets_good}. When t100 arrives at 9s, we examine the location updates in the past 10 seconds and find two officers u10 and u20 who recently updated their locations. We check the area codes of t100, u10, and u20 and produce a notification for u10. When t200 arrives at 28s, we look back in a 10-second window and find the location update from u10 at 22s, so we produce a notification for u10. Note that the location update from u20 at 13s is not ``used''. When t200 come at 28s, this location update of u20 is too old for the the tweet~\footnote{For illustrative simplicity, here we only look for location updates for new tweets. One may consider to look for tweets for new location updates and notify u20 about t100, but t200 for u20 would still be missing. Increasing the window size would work for this example but couldn't be applied for general cases.}.

\section{Experimental Results}
\label{sec:exprs}
In this section, we present a set of experiments conducted to evaluate the performance of the 
BAD system. We focused on the performance of BAD-CQ 
and compared that with the GOOD system described in Section~\ref{sec:good_vs_bad}. 
We first examined the basic ingestion and query performance of active datasets. 
Then, we investigated BAD-CQ's continuous channel performance regarding supportable subscribers in different use cases. 
Also, we compared the performance of the GOOD and the BAD systems using the same use cases.
Finally, we investigated the speed-up and scale-out performance of BAD-CQ when it is given more resources. Our experiments were conducted on a cluster connected using a Gigabit Ethernet switch (up to 16 nodes). Each node had a Dual-Core AMD Opteron Processor 2212 2.0 GHz, 8 GB of RAM, and a 900 GB hard disk drive.

\subsection{Active Dataset Scale-out Performance}
\begin{figure}[h]
    \centering
    \includegraphics[width=0.90\textwidth]{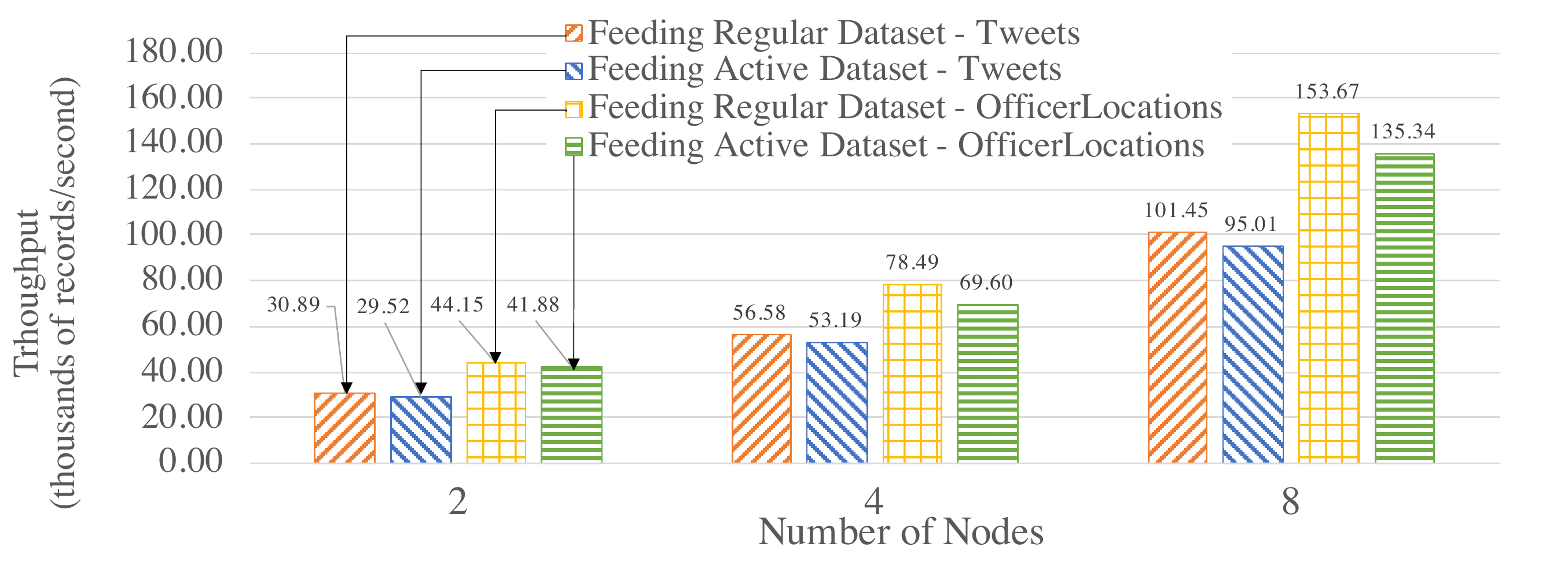}
    \caption{Ingestion performance on active datasets}
    \label{fig:expr_ing_perf}
\end{figure}

\begin{figure}[h]
    \centering
    \includegraphics[width=0.90\textwidth]{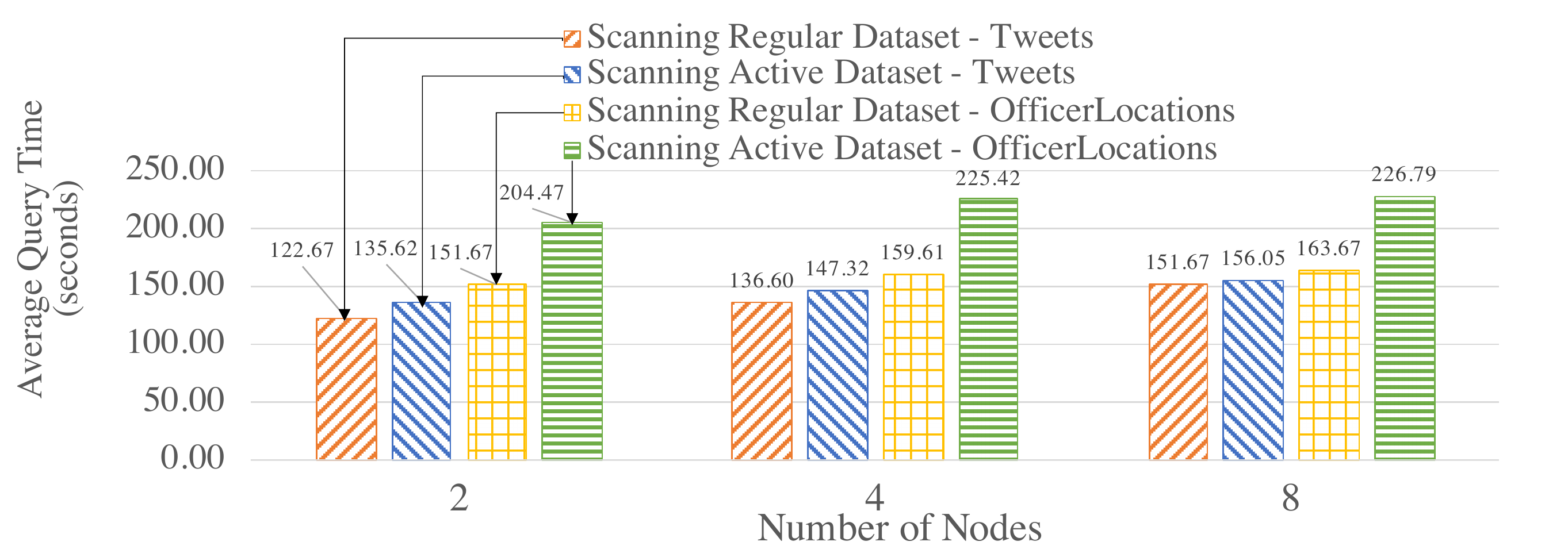}
    \caption{Query performance on active datasets}
    \label{fig:expr_scan_perf}
\end{figure}

Since active datasets store active timestamps with records for continuous channel evaluation, writing and reading active datasets will have the same additional cost due to the additional bytes. In order to examine the performance impact of that, we conducted ingestion and query performance experiments with active datasets. We used two types of data: the Tweets and OfficerLocations defined in Figure~\ref{ddl:tweets} and Figure~\ref{ddl:officer} respectively. Each tweet was around 140 bytes, and each user location record was around 60 bytes. An active timestamp was 9 bytes long (1 byte for data type and 8 bytes for epoch time). For both scale-out experiments, we started with 100 million records on a 2-node cluster and increased that to 400 million records on a 8-node cluster. For the ingestion performance experiments, we measured the ingestion throughput. For the query performance experiments, we measured the average time over 50 query executions for scanning all records in a dataset. The results are shown in Figure~\ref{fig:expr_ing_perf} and Figure~\ref{fig:expr_scan_perf} respectively.

When ingesting data into active datasets, the additional work comes from attaching active timestamps to incoming data records and persisting them into the storage engine. As we can see from Figure~\ref{fig:expr_ing_perf}, the ingestion throughput on both the Tweets and OfficerLocations datasets have some regression compared with the regular datasets. The throughput regression is proportional to the size ratio between an incoming record and the active timestamp. When an incoming record is big, the performance impact is relatively small and vice versa. With more nodes in the cluster, the throughput increases since more resources (CPU and storage bandwidth) can be used for parsing and storing incoming data.

When scanning active datatasets, the query time increases due to the additional cost of reading the large records with active timestamps from disk. Similarly, the query time increase is proportional to the size ratio between a stored record and the active timestamp. As the cluster size grows, the query time increases slightly due to the increased query execution cost on a larger cluster, but overall it remains stable since AsterixDB shards its stored data across all nodes.

\subsection{Channel Performance}
\label{sec:channel_expr}
As a channel runs periodically at a user specified period, it requires the channel evaluation to finish within that given period of time. The channel execution time depends on the channel query complexity and the size of the data involved (e.g., the number of tweets and subscribers). In order to examine the performance of data channels, we measured the maximum number of subscribers that can be supported by a channel within a given period. 
For these use cases, we introduce a new dataset \textbf{Schools}, defined in Figure~\ref{ddl:schools}, to store schools' information as relevant auxiliary information. A list of schools can be attached to hateful tweets to provide additional information for use by the responding in-field officers. The Schools dataset contains 10,000,000 records, and each record is around 70 bytes. We used the following four use cases to examine channel performance: 

\begin{enumerate}
    \item \textit{\textbf{NewLocalHatefulTweets}}: Send me new hateful tweets from a certain area (defined in Figure~\ref{ddl:new_local_tweets_chn}).
    \item \textit{\textbf{NewLocalHatefulTweetsWithSchools}}: Send me new hateful tweets from a certain area together with information about schools in that area (defined in Figure~\ref{ddl:new_local_tweets_with_schools_chn}).
    \item \textit{\textbf{NewNearbyHatefulTweets}}: Send me new hateful tweets nearby (defined in Figure~\ref{ddl:new_nearby_hateful_tweets_cq}).
    \item \textit{\textbf{UnseenNearbyHatefulTweets}}: Send me nearby hateful tweets that I've not seen before (defined in Figure~\ref{ddl:unseen_nearby_hateful_tweets_cq}).
\end{enumerate}

\begin{figure}[h]
\footnotesize
\begin{lstlisting}[
           language=SQL,
           basicstyle=\ttfamily,
           showstringspaces=false,
           commentstyle=\color{gray}
        ]
CREATE TYPE School AS OPEN {
  sid: int,
  area_code: string,
  name: string
};
CREATE DATASET Schools(School) PRIMARY KEY sid;
\end{lstlisting}
\caption{Datatype and dataset definition for Schools}
\label{ddl:schools}
\end{figure}

\begin{figure}[h]
\scriptsize
\begin{lstlisting}[
           language=SQL,
           basicstyle=\ttfamily,
           showstringspaces=false,
           commentstyle=\color{gray}
        ]
CREATE CONTINUOUS CHANNEL NewLocalHatefulTweets(area_code) PERIOD duration("PT10S") {
    SELECT t FROM Tweets t 
    WHERE t.area_code = area_code AND is_new(t)
};
\end{lstlisting}
\caption{A continuous channel for new local hateful tweets}
\label{ddl:new_local_tweets_chn}
\end{figure}

\begin{figure}[h]
\scriptsize
\begin{lstlisting}[
           language=SQL,
           basicstyle=\ttfamily,
           showstringspaces=false,
           commentstyle=\color{gray}
        ]
  CREATE CONTINUOUS CHANNEL NewLocalHatefulTweetsWithSchools(area_code) 
   PERIOD duration("PT10S") {
    SELECT t, 
    (SELECT VALUE s FROM Schools s WHERE s.area_code = t.area_code) AS nearby_schools
    FROM Tweets t 
    WHERE t.area_code = area_code AND is_new(t)
  };
\end{lstlisting}
\caption{A continuous channel for new local hateful tweets with schools}
\label{ddl:new_local_tweets_with_schools_chn}
\end{figure}

In use cases 1 and 2, subscribers subscribe to a channel with their interested area codes. In use cases 3 and 4, subscribers subscribe with their officer IDs and their locations are spatial data mapped to IDs. All channels were configured to execute every 10 seconds. To approximate incoming data in practice, we set up external programs that sent tweets and officer location updates continuously. For tweets, the client program sent at a configurable rate (tweets / second), and 10\% of the incoming tweets were hateful. For location updates, the client program sent location updates on behalf all subscribers (in-field officers), and an average of 1/3 of the in-field officers updated their locations every 10 seconds. Both programs ran on machines outside of the BAD cluster.

\begin{figure}[h]
    \centering
    \includegraphics[width=0.92\textwidth]{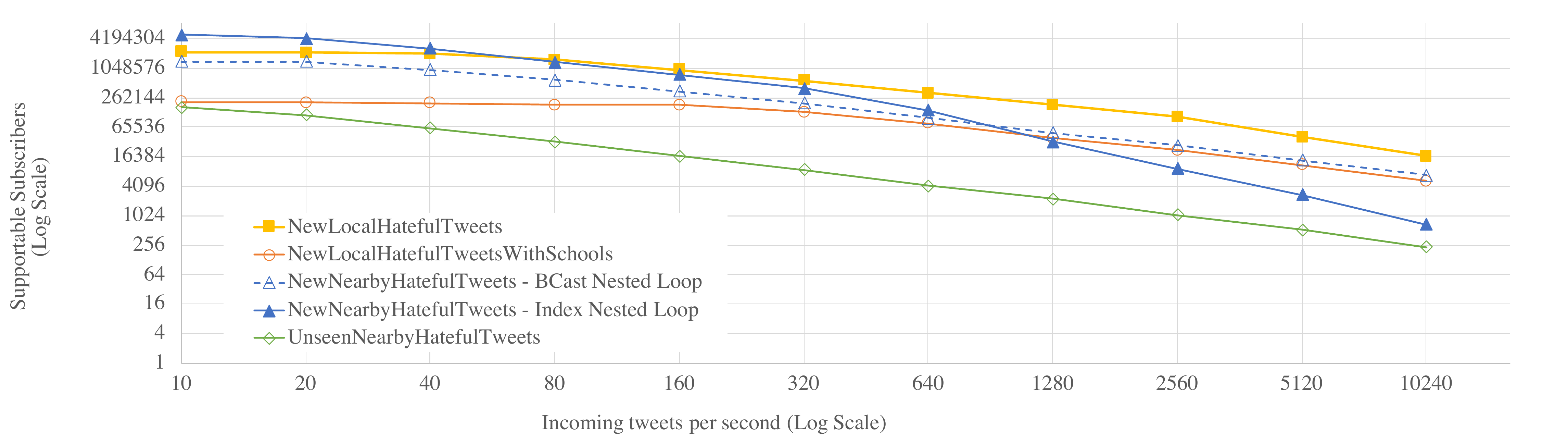}
    \caption{Maximum number of supportable subscribers under different incoming data rates}
    \label{fig:channel_perf}
\end{figure}

In all four use cases, we fixed the incoming tweet rate and searched for the maximum number of supportable subscribers in the given 10-second channel execution period while both tweets and location updates were coming. We varied the incoming tweet rate to see how channel performance changed. For the ``NewNearbyHatefulTweets'' channel in particular, we chose two algorithms (broadcast nested loop join and index nested loop join) to evaluate the spatial join between the incoming tweets and officers' locations. (We broadcast data from the Tweets dataset and utilized the R-Tree index on the location attribute of the OfficerLocations dataset.) We deployed BAD-CQ on a 6-node cluster and the performance results are shown in Figure~\ref{fig:channel_perf}. (Note the use of a log scale for the y-axis.)

Depending on the channel query complexity, the maximum number of supportable subscribers varies. For all four use cases, the maximum number of supportable subscribers decreases as the incoming tweet rate increases; this is due to the increased cost of producing and persisting~\footnote{As mentioned in Section~\ref{sec:bad_data_delivery}, BAD persists customized data to disk by default to allow brokers to pull later.} more customized notifications. Comparing the results for ``NewLocationHatefulTweets'' and ``NewLocalHatefulTweetsWithSchools'', we see that the latter one has lower performance, as adding in school information incurs more computational and persistence cost. Comparing ``NewNearbyHatefulTweets - BCast Nested Loop'' and ``NewNearbyHatefulTweets - Index Nested Loop'', we see that the use of the index offers much better performance than scanning the whole OfficerLocations dataset when the incoming tweet rate is low. As the incoming tweet rate grows, however, the performance of the index nested loop join becomes worse than the broadcast join. The reason is that, with more incoming tweets, the maximum number of supportable subscribers decreases due to the increased cost of computing customized data. For the join operation between tweets and officer locations, then, having more tweets and fewer actual subscribers (in-field officers) increases the query's selectivity for OfficerLocations. Since the index nested loop join accesses the primary index through a secondary index, when the selectivity becomes high, the performance of using that index becomes worse than just scanning the primary dataset. Interested readers may refer to \cite{luoc_lsm} for a more detailed analysis of the underlying storage engine's performance benchmarks.

\subsection{Good vs. BAD Performance}
The BAD system enables developers to create BAD services with declarative statements. The GOOD system, in contrast, requires developers to manually glue multiple systems together and orchestrate them programmatically to create BAD services. In order to show that the BAD system not only alleviates developers' effort when creating BAD services, but can also provide better performance compared with a GOOD system, we chose several use cases supported by both the BAD and GOOD systems and measured their performance on both.

We used the GOOD system detailed in Section~\ref{sec:a_good_system} for these experiments. As we discussed in Section~\ref{sec:good_vs_bad}, the GOOD system cannot provide all query semantics supported in the BAD system. Not all use cases in Section~\ref{sec:channel_expr} can be supported directly in the GOOD system. Spark Structured Streaming does not support spatial joins between streams, so we used area code to represent the location of tweets and officers. The use cases used for comparing the performance of the BAD system and the GOOD system are as follows:

\begin{enumerate}
    \item \textbf{\textit{NewLocalHatefulTweets}}: Send me new hateful tweets from a certain area (same as Section~\ref{sec:channel_expr}). 
    \item \textbf{\textit{NewLocalHatefulTweetsWithSchools}}: Send me new hateful tweets from a certain area together with the schools in that area (same as Section~\ref{sec:channel_expr}).
    \item \textbf{\textit{NewHatefulTweetsForLocalActiveUsers}}: Send me new hateful tweets from the same area as my current location (similar to NewNearbyHatefulTweets in Section~\ref{sec:channel_expr}, but modified to use area\_code for this experiment).
\end{enumerate}

\begin{figure}[h]
    \centering
    \includegraphics[width=0.95\textwidth]{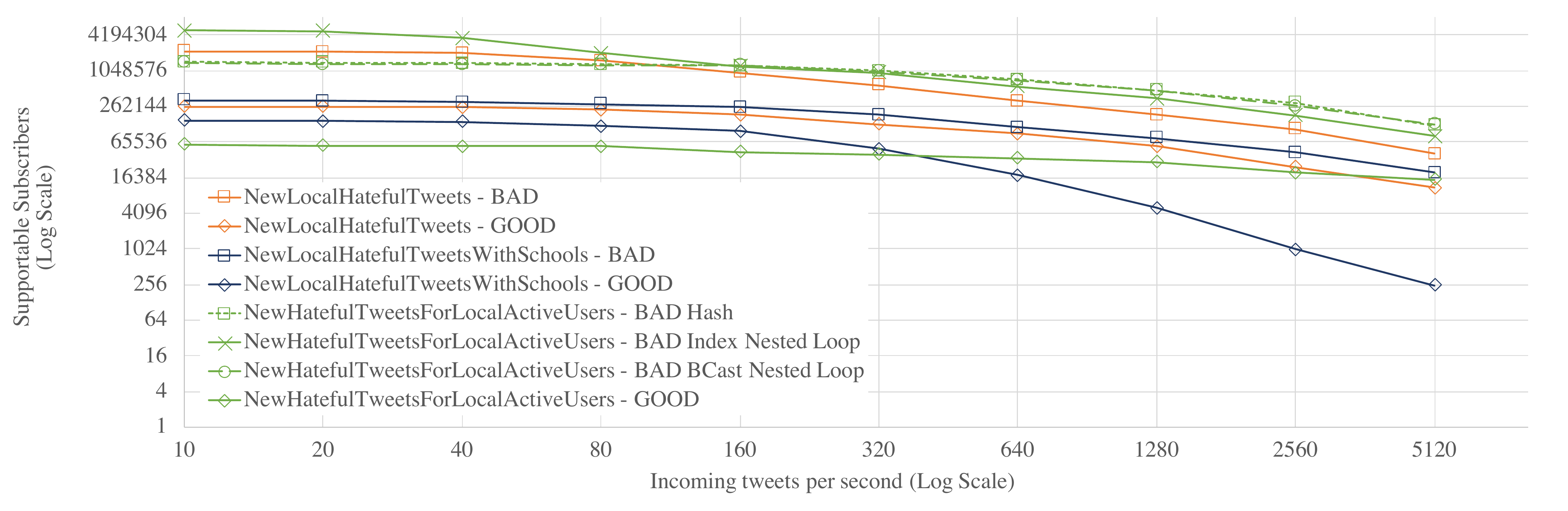}
    \caption{Performance comparison of BAD-CQ and the GOOD system}
    \label{fig:expr_good_perf}
\end{figure}

In use cases 1 and 2, subscribers subscribe to a channel with the area codes of interest. In use case 3, subscribers subscribe to the channel with the their officer IDs. Due to the high overhead of integrating Spark Structured Streaming with MongoDB, we tuned down the size of the \textbf{Schools} dataset by 5x to 2,000,000. To demonstrate the advantages that the BAD system as of utilizing indexes and different query evaluation algorithms, we picked the ``NewHatefulTweetsForLocalActiveUsers'' use case, and we experimented with hash join, broadcast nested loop join, and index nested loop join. In this experiment, we focused on the processing core of both systems without including result delivery using brokers. The generated notifications were persisted in storage, as in the default pull mode. All incoming data was persisted as well for retrospective analysis. The performance results in terms of the number of supportable subscribers are shown in Figure~\ref{fig:expr_good_perf}. (Note the use of a log scale for the y-axis.)

In all three cases, BAD-CQ outperforms the GOOD system. As the incoming tweet rate grows, the performance of both systems drop because of the increased cost of producing and persisting more notifications. Similar to Section~\ref{sec:channel_expr}, both systems have better performance for ``NewLocalHatefulTweets'' (colored in orange) than for ``NewLocalHatefulTweetsWithSchools'' (colored in blue) due to the additional cost of attaching relevant school information. In particular, the ``NewLocalHatefulTweetsWithSchools'' use case for GOOD suffers more from the increased incoming tweet rate, as the cost of persisting notifications with schools into MongoDB becomes high when there are many notifications. For ``NewHatefulTweetsForLocalActiveUsers'', we see a similar performance benefit for utilizing an index and the same performance regression when the incoming tweet rate becomes high. Hash join offered only a slight advantage over a broadcast nested loop join in this case, as the total number of tweets for each channel execution is relatively small.

\begin{figure}[h]
    \centering
    \includegraphics[width=0.60\textwidth]{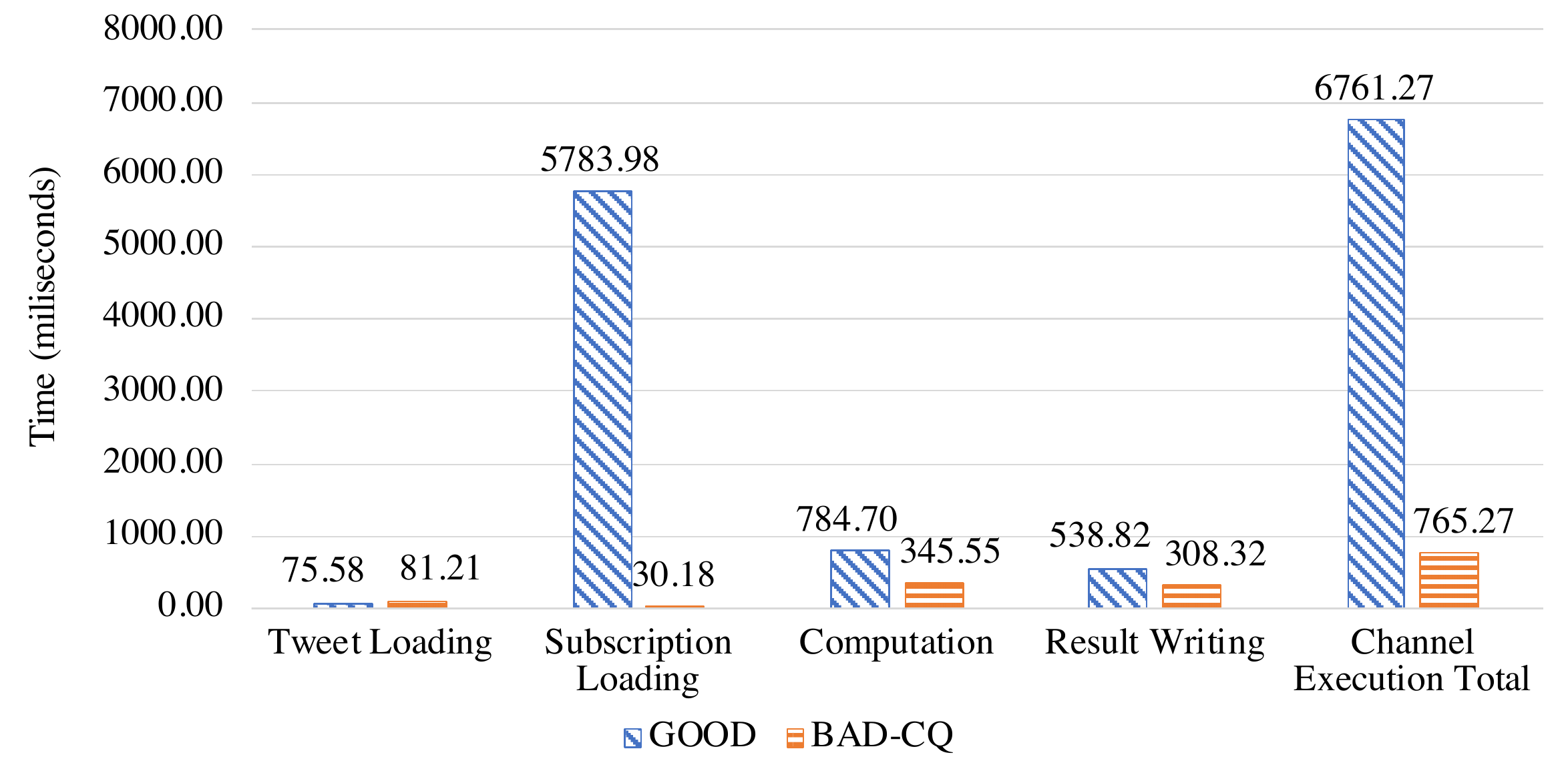}
    \caption{Cost of ``NewLocalHatefulTweets'' with 150,000 subscribers and 80 tweets/second on both the GOOD and BAD system}
    \label{fig:good_breakdown}
\end{figure}

In order to better understand the cost of the GOOD system, we chose the ``NewLocalHatefulTweets'' use case with 150,000 subscribers and 80 tweets/second and measured the time consumed by each stage of its channel execution on both the GOOD and BAD system. The result is shown in Figure~\ref{fig:good_breakdown}, which also includes the overall channel execution time.
As can be seen, the GOOD channel execution spent much of its time loading Subscriptions from MongoDB. This is a consequence of the overhead of gluing different systems together, as shipping data from one sub-system to another incurs additional serialization/deserialization and data transformation and transmission costs. One could consider maintaining copies of the relevant data and subscriptions in Spark Structured Streaming as well, to accelerate the processing, but then developers would have to handle consistency challenges and need to migrate updates back and forth between Spark Structured Streaming and MongoDB. In contrast, BAD-CQ spent much less time on subscription loading. Since tweets were being ingested at the same time, there was a bit of read/write contention on the \textit{Tweets} dataset that caused the tweet loading time to be higher than the subscription loading time on BAD-CQ.

\subsection{BAD Scalability}

Finally, we investigated the scalabilty of BAD-CQ from two angles: \textit{speed-up} - given a fixed workload, see if the performance improves with more resources, and \textit{scale-out} - increase the workload together with available resources to see if the performance remains stable. We chose the ``NewNearbyHatefulTweets - Bcast Nested Loop'' channel and increased the channel's period to 30 seconds for this experiment. All other settings were the same as Section~\ref{sec:channel_expr}.

\begin{figure}[h]
    \centering
    \includegraphics[width=0.50\textwidth]{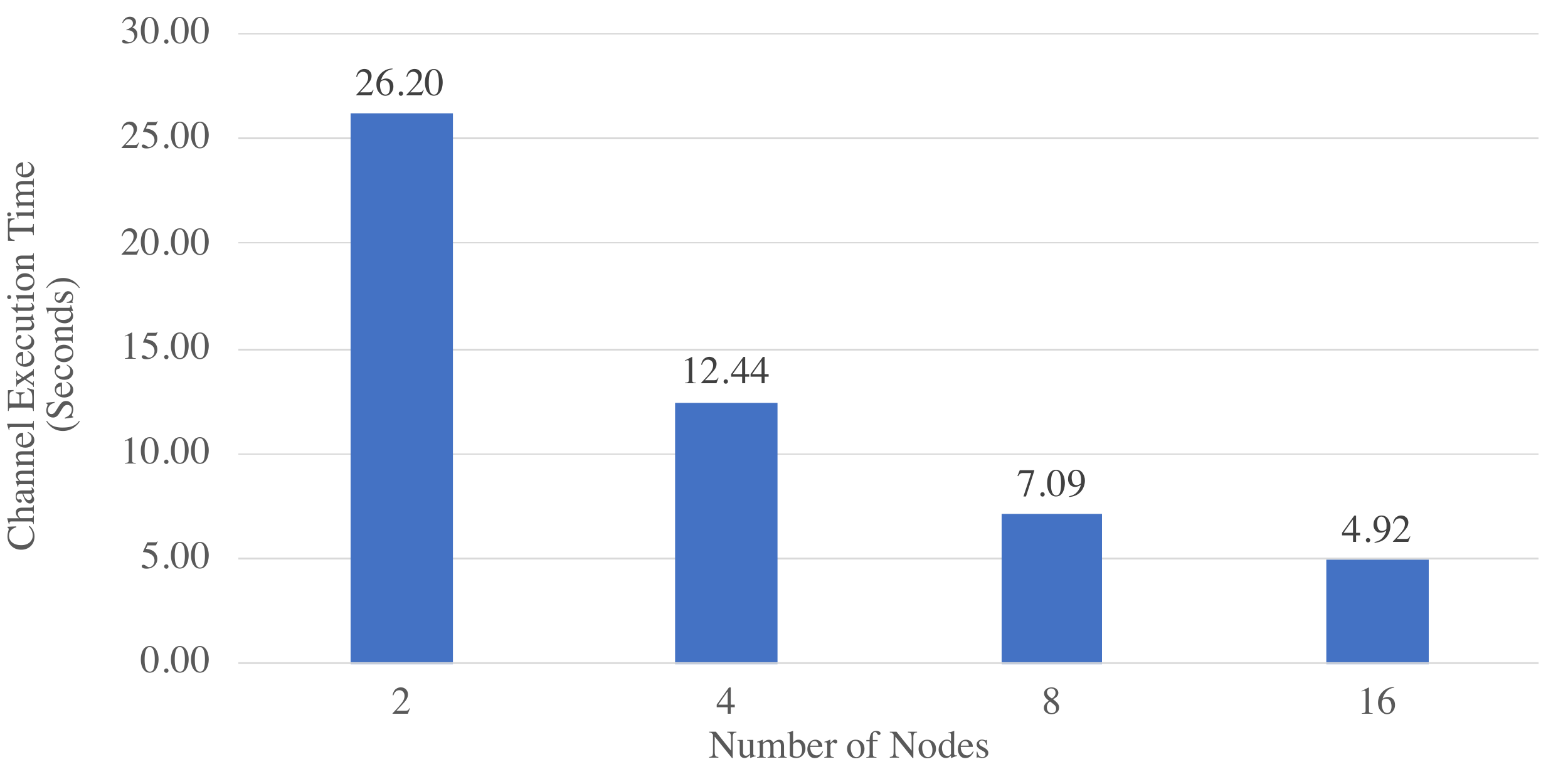}
    \caption{Speed-up BAD-CQ with fixed incoming tweet rate and number of subscribers}
    \label{fig:speed_up}
\end{figure}

\textbf{Speed-up experiments:} The channel workload is determined by the incoming tweets per second and the number of subscribers (in-field officers). In this experiment, we fixed the incoming tweet rate to 160 tweets per second and had 140,000 subscribers. We increased the cluster size from 2 nodes to 4, 8, and 16 nodes, and we measured the channel execution times, as shown in Figure~\ref{fig:speed_up}. When the cluster grows, the channel execution time is almost halved because the subscribers' locations are stored on twice as many machines. Since each node now has less data, the join between incoming tweets and officer's locations, which computes on all nodes, can finish sooner. As tweets are broadcast to all nodes in the cluster and the execution overhead also grows with the cluster size, the speed-up gain gradually decreases with larger cluster sizes.

\begin{figure}[h]
    \centering
    \includegraphics[width=0.50\textwidth]{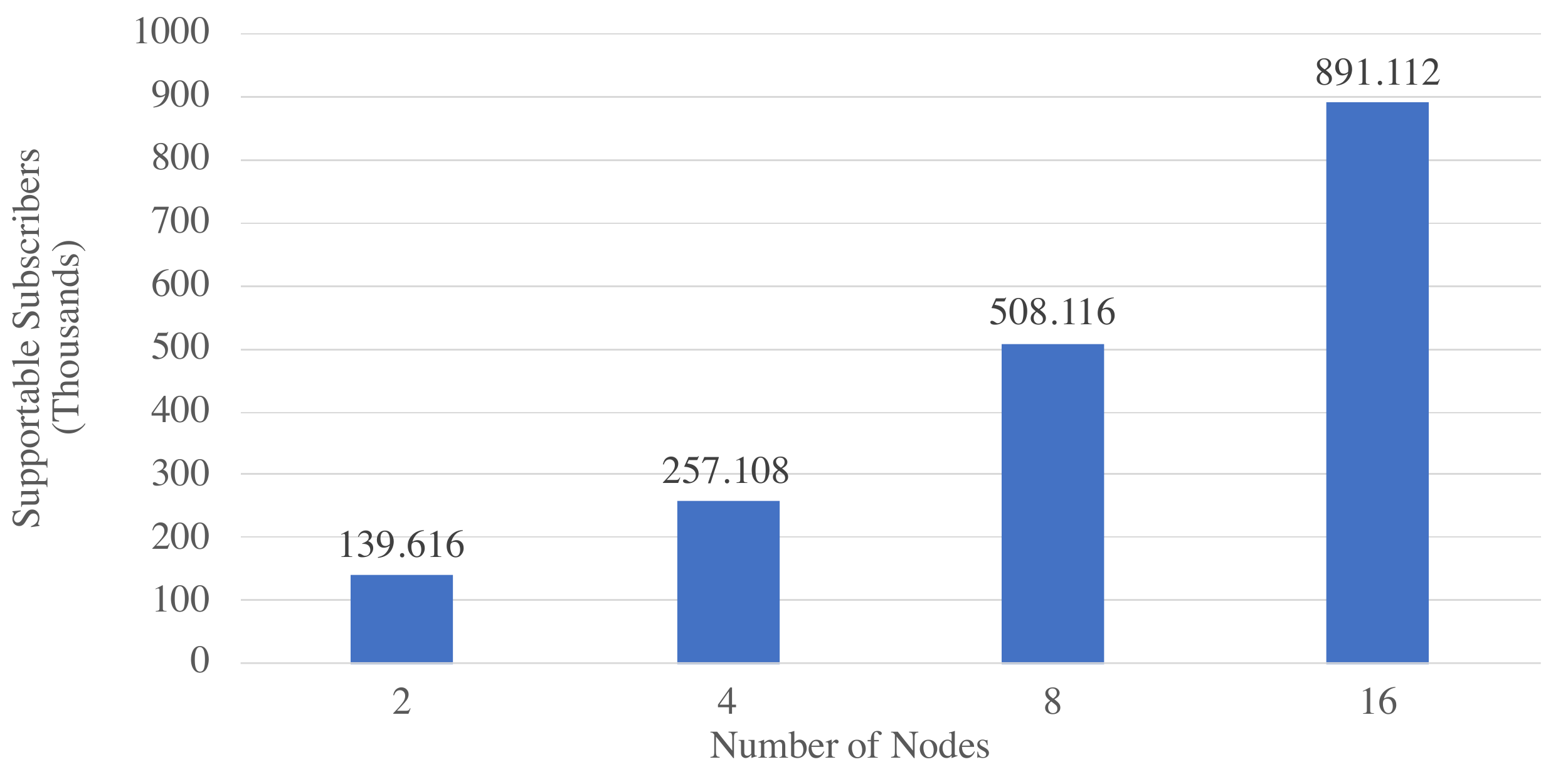}
    \caption{Scale-out BAD-CQ with fixed incoming tweet rate}
    \label{fig:scale_out_fix_tweets}
\end{figure}

\textbf{Scale-out experiments:} We used two experiments to evaluate the scale-out performance of BAD-CQ. We first fixed the incoming tweet rate to 160 tweets/second and increased the cluster size from 2 nodes to 4, 8, and 16 nodes to see how many subscribers could be supported in each setting. The result is shown in Figure~\ref{fig:scale_out_fix_tweets}. As we double the size of the cluster, the maximum number of supportable subscribers almost doubles. 
Similar to the speed-up experiment, twice many nodes allow the join operation to handle more data in the given time period. 

\begin{figure}[h]
    \centering
    \includegraphics[width=0.50\textwidth]{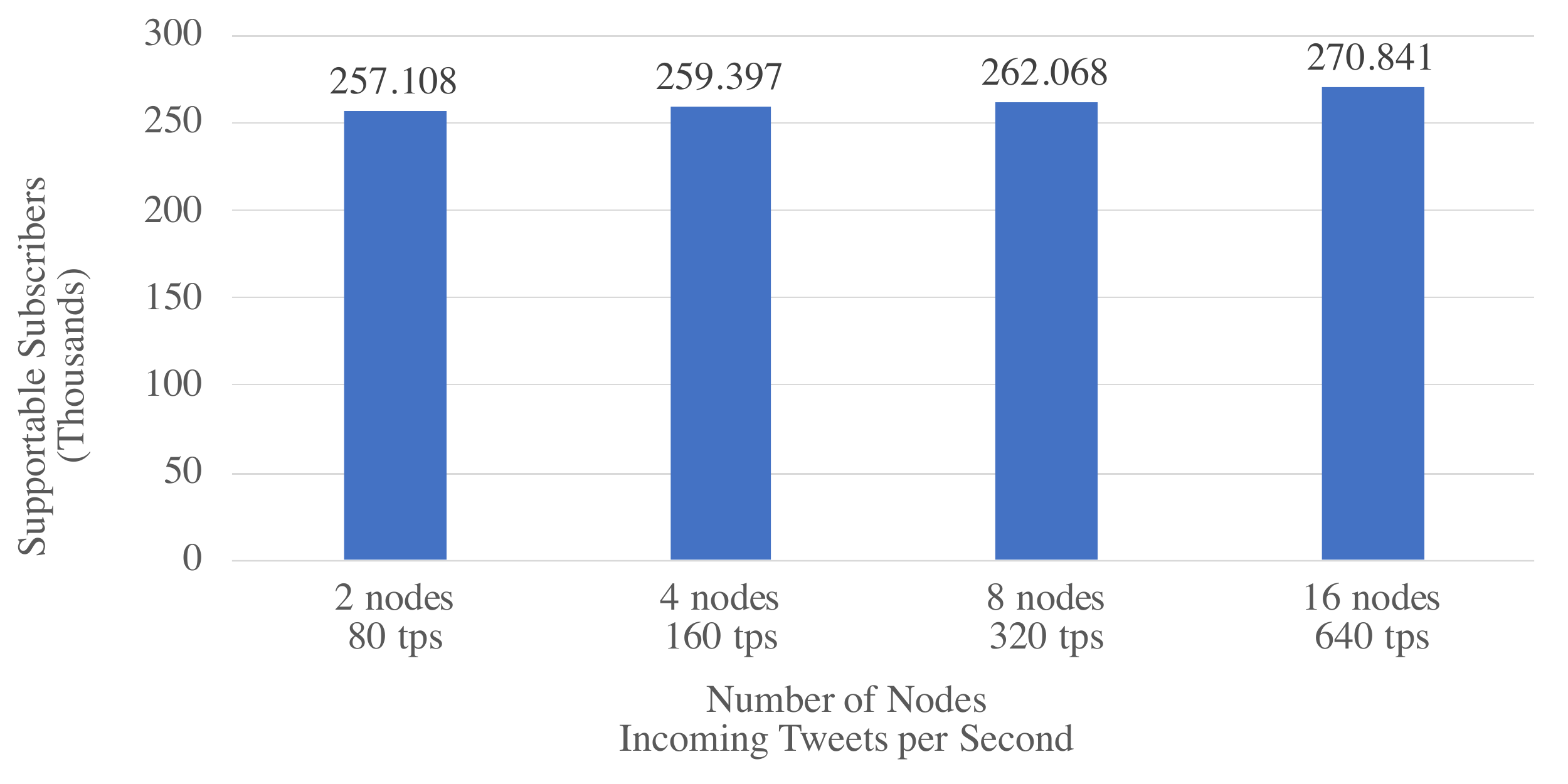}
    \caption{Scale-out BAD-CQ with increasing incoming tweet rate}
    \label{fig:scale_out_chaing_tweets}
\end{figure}

In the second experiment, we increased the incoming tweet rate together with the cluster size. We started with a 2-node cluster with 80 incoming tweets per second, and we increased the cluster size and the incoming tweet rate by the factor of two, up to 16 nodes and 640 tweets per second. The result is shown in Figure~\ref{fig:scale_out_chaing_tweets}. The channel performance maintains relatively stable as we increase the workload and add more resources at the same time.

\section{Conclusions and Future Work}
In this work, we considered a world where Big Data is no longer just bytes sitting on storage devices, waiting to be analyzed, but is valuable information surrounded by active requests asking for continual ``news updates''. In such a Big Active Data (BAD) world, developers often need to create and manage data services to support analysts in working with declarative queries and subscribers looking for the latest updates. In order to reduce the effort for developers creating BAD services, we have built the BAD system, consisting of BAD-RQ, which ``activates'' a parameterized query as a data channel for subscribers to receive periodic query results of interest, and BAD-CQ, which introduces continuous (incremental query) semantics into data channels and optimizes the channel infrastructure for continuous use cases. 
We showed the user model, design, and implementation of our system and illustrated how developers can use it to create BAD services declaratively. To demonstrate the complexity of creating BAD services without BAD, we also presented a ``GOOD'' system created by gluing multiple Big Data systems together. We examined the performance of the BAD system under different workloads and compared that with an instance of a GOOD system. The results for the use cases examined showed that the BAD system could support up to four millions subscribers on a six-node cluster, was able to horizontally scale out with more resources, and offered significantly better performance as compared with the GOOD system. In all, the BAD system provides a systematic solution for creating BAD services at scale. 

This work leads to a number of interesting opportunities for future investigation:

\begin{itemize}
    \item \textbf{Connecting multiple BAD systems:} In a BAD world, there could be multiple BAD systems running and managed by different organizations. In some use cases, developers may need to share information between different organizations and combine it with local data to create applications.
    Building a scalable data sharing service from scratch requires a lot of work. With BAD, we could allow developers to connect multiple BAD systems via data channels and feeds. Developers of multi-site applications could then benefit from the BAD approach and could create data sharing services with very little implementation and management overhead.
    \item \textbf{Exploiting shared computation among data channels:} In the current BAD system, data channel queries are processed, compiled, and optimized independently. While shared computation arises from evaluating the parameterized requests within a given channel together, more exploitation of sharing is possible. 
    Similar to \cite{niagara_cq}, we could analyze multiple data channel queries, split them into smaller parts, discover shared computations, and reuse intermediate results to improve channel performance by avoiding redundant computation.
    \item \textbf{Resource management \& scheduling of channel executions:} 
    Currently, every channel execution is scheduled based on its period. Each channel execution runs as an independent job in the analytical engine, and an internal resource manager manages the resource usage of all jobs running in the system. When there is resource contention, certain channel executions may be delayed and cause a channel to terminate (as we require channel executions to finish within the given period to meet the channel's time requirement). Given different channel periods and users' quality of service requirements, it should be possible to develop a smarter scheduling strategy in which we allow more flexible channel execution schedules based on the available resources and obtain better resource utilization at the same time.
\end{itemize}

\begin{acks}
This research was partially supported by NSF grants IIS-1447826, IIS-1447720, IIS-1838222, IIS-1838248, CNS-1924694 and CNS-1925610.
\end{acks}

\bibliographystyle{ACM-Reference-Format}
\bibliography{sample-manuscript}



\end{document}